\documentclass{aa}
\usepackage[varg]{txfonts}

\begin{document}

   \title{The fate of planetesimals formed at planetary gap edges}

   \author{Linn E.J. Eriksson
          \inst{1}
          ,
          Thomas Ronnet \inst{1}
          , 
          Anders Johansen\inst{1,2}
          }

   \authorrunning{L.E.J. Eriksson}
   \institute{Lund Observatory, Department of Astronomy and Theoretical Physics, Lund University, Box 43, 221 00 Lund, Sweden \and Center for Star and Planet Formation, GLOBE Institute, University of Copenhagen, \O ster Voldgade 5-7, 1350 Copenhagen, Denmark\\
              \email{linn@astro.lu.se}}

   \date{Received X; accepted X}

\abstract{ The presence of rings and gaps in protoplanetary discs are often ascribed to planet-disc interactions, where dust and pebbles are trapped at the edges of planetary induced gas gaps. Recent work has shown that these are likely sites for planetesimal formation via the streaming instability. Given the large amount of planetesimals that potentially form at gap edges, we address the question of their fate and ability to radially transport solids in protoplanetary discs. We perform a series of N-body simulations of planetesimal orbits, taking into account the effect of gas drag and mass loss via ablation. We consider two planetary systems: one akin to the young Solar System, and another one inspired by the structures observed in the protoplanetary disc around HL Tau. In both systems, the close proximity to the gap-opening planets results in large orbital excitations, causing the planetesimals to leave their birth locations and spread out across the disc soon after formation. We find that collisions between pairs of planetesimals are rare, and should not affect the outcome of our simulations. Collisions with planets occur for $\sim\!\! 1\%$ of the planetesimals in the Solar System and for $\sim\!\! 20\%$ of the planetesimals in the HL Tau system. Planetesimals that end up on eccentric orbits interior of $\sim\!\!  10\, \textrm{au}$ experience efficient ablation, and lose all mass before they reach the innermost disc region. In our nominal Solar System simulation with a stellar gas accretion rate of $\dot{M}_0=10^{-7}\, M_{\odot}\, \textrm{yr}^{-1}$ and $\alpha=10^{-2}$, we find that 70\% of the initial planetesimal mass has been ablated after $500\, \textrm{kyr}$. Since the protoplanets are located further away from the star in the HL Tau system the ablation rate is lower, and only 11\% of the initial planetesimal mass has been ablated after $1\, \textrm{Myr}$ using the same disc parameters. The ablated material consist of a mixture of solid grains and vaporized ices, where a large fraction of the vaporized ices re-condense to form solid ice. Assuming that the solid grains and ices grow to pebbles in the disc midplane, this results in a pebble flux of $\sim\!\! 10-100\, M_{\oplus}\, \textrm{Myr}^{-1}$ through the inner disc. This happened at a time in the Solar System so early that there is likely no record of it. Our results demonstrate that scattered planetesimals can carry a significant flux of solids past planetary-induced gaps in young and massive protoplanetary discs.  } 

\maketitle 

\section{Introduction}
Concentric rings and gaps in the millimeter continuum emission associated with pebbles are commonly observed features in protoplanetary discs (e.g. \citealt{ALMA2015,Andrews2018}). Recent evidence suggests that at least some of these radial pebble concentrations are due to trapping at pressure bumps \citep{Dullemond2018}. The origin of pressure bumps is often ascribed to planet-disc interactions, where a growing protoplanet carves a gap in the gas disc, leading to the formation of a pressure maximum at the planetary gap edge (e.g. \citealt{Pinilla2012,Dipierro2015,Fedele2018,Zhang2018,Favre2019}). The direct observation of protoplanets orbiting within protoplanetary disc gaps support this scenario (\citealt{Pinte2019,Pinte2020}). Since the ratio of solids-to-gas in these pressure bumps can grow to values significantly higher than the global one, they are favorable spots for planetesimal formation via the streaming instability (SI) (\citealt{YoudinGoodman2005,Johansen2009,Lyra2009,BaiStone2010,Carrera2015,Yang2017}). 

In \citet{Eriksson2020} (hereafter referred to as Paper 1), we performed global 1D simulations of dust evolution and planetesimal formation via the SI in a protoplanetary disc with multiple gap-opening planets. We found that planetesimals form at the gap edges for a wide range of planetary masses, particle sizes and disc parameters. A similar study by \citet{Stammler2019} show that planetesimal formation via the streaming instability can further explain the observed range of optical depths among dust rings in the DSHARP survey.  Results from the first 3D simulations of planetesimal formation in a pressure bump were reported in \citet{Carrera2021}, who concluded that planetesimal formation in pressure bumps is a robust process in the case of cm-sized particles. They did however not find any planetesimal formation in the case of mm-sized particles, but higher resolution and a broader parameter search is needed to assess this.

The above studies tell us that planetesimal formation in pressure bumps likely is a common process. In this work we assume that these pressure bumps are formed by growing planets, and we ask ourselves the question: what is the fate of planetesimals formed at these planetary gap edges? To answer this question we perform a suite of N-body simulations, taking into account the effect of gas drag on the planetesimals and mass loss via ablation. We consider two planetary systems: the Solar System with Jupiter and Saturn and an HL Tau inspired system with 3 planets. The main points we want to address are: (1) to what extent do gravitational interactions with the forming planets redistribute the planetesimals formed at their gap edge?; (2) can the frictional heating of planetesimals on eccentric orbits drive the production of a significant amount of dust through surface ablation?; and (3) how common are collisions between pairs of planetesimals, and between planetesimals and planets? 

We find that planetesimals which form at the edges of planetary gaps do not remain at their birth location. Gravitational scattering by the embedded planets cause the planetesimals to spread out across the disc and result in high orbital eccentricities. If the planetesimals end up in the inner part of the disc, efficient ablation due to high surface temperatures results in fast mass loss; this prevents any planetesimal from entering the innermost disc region. Collisions between pairs of planetesimals are rare and should not affect the outcome of our simulations, whereas planet-planetesimal collisions do occur, and more often so in the HL Tau system than in the Solar System. In general, the closer proximity to the Sun results in much more ablation in the Solar System than in the HL Tau system. If this ablated material re-condenses to form pebbles in the disc midplane, the result is a significant flux of pebbles both interior and exterior of Jupiter's orbit. Since the ablation efficiency depends strongly on the gas density, this pebble flux should only be present in relatively young and massive discs.

In Sect. \ref{sec:theory}, we present our disc model, our prescriptions for gas drag and mass ablation, and our calculations of condensation temperatures and planetsimal-planetesimal collision timescsales. The numerical setup and initial conditions for our simulations are presented in Sect. \ref{sect: numerical methods}. Results from the Solar System simulations are presented in Sect. \ref{sect: solar system}, and results from the HL Tau simulations are presented in Sect. \ref{sect:HL Tau}. In Sect. \ref{sec: the ablated material} we calculate the pebble flux that derives from planetesimal ablation and discuss how this fits into our understanding of the isotope dichotomy in the Solar System. The effect of planet-planetesimal collisions are discussed in Sect. \ref{sect: collisions}. In Sect. \ref{setc: conclusion}, we summarize the key findings of the paper. Further information about the the ablation model and some additional figures can be found in Appendix \ref{Appendix: noAbl}-\ref{Appendix: 6plan}.

\section{Theory}\label{sec:theory}
The protoplanetary disc gas was modeled using a static 1-D accretion disc. We added an acceleration to the equation of motion of the planetesimals in order to account for the friction force they feel as they move through the gas. This friction acts to heat up the planetesimals and causes vaporization of the surface layers. We calculate the rate at which material is ablated from the planetesimals and update their mass at each timestep in the simulation. The temperature and mass loss rate are highly dependent on the composition of the planetesimals, which is set by their formation location relative to the location of the major icelines in the disc. We consider the disc to contain the following volatiles -- H$_2$O, CO$_2$ and CO -- and calculate their respective condensation temperatures. Finally, we estimate the frequency of planetesimal-planetesimal collisions using an analytic expression.

\subsection{Disc model}
\subsubsection*{Surface density $\Sigma$}
The gas surface density profile was defined to be that of a 1-D steady disc,
\begin{equation}\label{eq: sigma0}
\Sigma=\frac{\dot{M}_0}{3\pi \nu} \exp \left[-\frac{r}{r_\textrm{out}}\right],
\end{equation}
where $\Sigma$ is the gas surface density, $\dot{M}_0$ is the disc accretion rate, $\nu$ is the kinematic viscosity of the disc, $r$ is the semimajor axis, and $r_{\rm out}$ is the position of the outer disc edge (e.g., \citealt{Pringle1981}). We used the alpha approach from \citet{ShakuraSunyaev1973} to approximate the kinematic viscosity, 
\begin{equation}
\nu = \alpha \Omega H^2,
\end{equation}
where $\alpha$ is a parameter that determines the efficiency of viscous transport, $\Omega  = (GM_*/r^3)^{1/2}$ is the Keplerian angular velocity, and $H$ is the scale height of the gas disc. We calculated the scale height as
\begin{equation}
H = c_{\textrm{s}}/\Omega, 
\end{equation}
where $c_s$ is the sound speed,
\begin{equation}\label{eq:cs}
c_{\textrm{s}} = \left(\frac{k_{\textrm{B}} T}{\mu m_{\textrm{H}}}\right)^{1/2}.
\end{equation}
In the equation above, $k_{\textrm{B}}$ is the Bolzmann constant, $T$ is the temperature, $\mu$ is the mean molecular weight, and $m_{\textrm{H}}$ is the mass of the hydrogen atom. The mean molecular weight was set to be 2.34, corresponding to a solar-composition mixture of hydrogen and helium \citep{Hayashi1981}. The midplane temperature of the disc was approximated using a fixed powerlaw structure, 
\begin{equation}\label{eq:Tdisc}
T=150\, \textrm{K} \times (r/{\rm AU})^{-3/7},
\end{equation}
where $150\, \textrm{K}$ is the temperature at $1\, \textrm{AU}$ \citep{ChiangGoldreich1997}.

\subsubsection*{Planetary gaps}
Massive planets push away material from the vicinity of their orbits, and as a result opens up a gap in the disc. We modeled these planetary gaps using a simple approach with Gaussian gap profiles. The Gaussian is described by the equation
\begin{equation}
G(r) = \frac{\Sigma_{a,0}}{\Sigma_{a,\textrm{min}}} \exp \left[-\frac{(r-a)^2}{2H_a^2}\right],
\end{equation}
where $\Sigma_{a,0}$ is the unperturbed surface density at the location of the planet, $\Sigma_{a,\textrm{min}}$ is the surface density at the bottom of the gap, and $a$ is the semimajor axis of the planetary orbit.  The depth of the gap is calculated as
\begin{equation}
\frac{\Sigma_{a,\textrm{min}}}{\Sigma_{a,0}} = \frac{1}{1+0.04K},
\end{equation}
where $K$ is given by
\begin{equation}
K=q^2\left(\frac{H_a}{a}\right)^{-5}\alpha^{-1},
\end{equation}
and $q$ is the planet to star mass ratio (e.g., \citealt{Kanagawa2015}). Each planet contributes their own Gaussian, and the final surface density profile is then obtained by dividing equation \ref{eq: sigma0} with $1+G(r)_{\textrm{1}}+G(r)_{\textrm{2}}+...$

\subsubsection*{Gas density $\rho$}
We use the following expression to obtain the midplane gas density,
\begin{equation}\label{eq:rhog0}
\rho(z=0) = \frac{\Sigma}{\sqrt{2\pi}H}.
\end{equation} 
In order to find the gas density at some height $z$ away from the midplane, we assume vertical hydrostatic equilibrium for the gas in the disc. We then end up with the equation
\begin{equation}
\rho(z) = \rho(z=0) \exp \left[\frac{-z^2}{2H^2}\right].
\end{equation}

\subsection{Drag force}
Planetesimals which are moving through the disc at a velocity different from that of the gas feel either headwind or a tailwind, and as a result they are decelerated or accelerated towards the value of the gas velocity. This change in velocity occurs on a timescale called the stopping time, $t_s$. Assuming that the planetesimals are spherical, their stopping time is
\begin{equation}
t_s = \left(\frac{\rho}{\rho_{\bullet}}\frac{v_{\textrm{th}}}{R_{\textrm{pl}}} \min \left[1,\frac{3}{8}\frac{v_{\textrm{rel}}}{v_{\textrm{th}}}C_D(Re)\right]\right)^{-1},
\end{equation}
where $\rho_{\bullet}$ and $R_{\textrm{pl}}$ are the solid density and radius of the planetesimals, $v_{\textrm{rel}}$ is the relative velocity between the planetesimals and the gas, $v_{\textrm{th}} = \sqrt{8/\pi}c_s$ is the gas thermal velocity, $C_D$ is the dimensionless drag coefficient, and $Re$ is the Reynolds number (e.g., \citealt{PeretsMurrayClay2011,Guillot2014}). The dimensionless drag coefficient is calculated as
\begin{equation}
C_D = \frac{24}{Re}(1+0.27Re)^{0.43}+0.47(1-\exp[-0.04Re^{0.38}])
\end{equation}
\citep{PeretsMurrayClay2011}, and the Reynolds number is
\begin{equation}
Re = \frac{4R_{\textrm{pl}}v_{\textrm{rel}}}{c_s l_g},
\end{equation}
where $l_g \sim 5\times 10^{-6}\, \textrm{kg}\, \textrm{m}^{-3} / \rho$ is the mean-free path of the gas \citep{SupulverLin2000}.

The acceleration of the planetesimal due to gas drag is calculated as
\begin{equation}
\textbf{\textit{a}}_{\textrm{drag}} = -\frac{1}{t_s}(\textbf{\textit{v}}_{\textrm{pl}}-\textbf{\textit{v}}_{\textrm{gas}}),
\end{equation}
where $\textbf{\textit{v}}$ is a velocity vector (e.g., \citealt{Whipple1972}). We assume that the gas velocity in the $z$-direction is zero, and obtain the Cartesian components of the gas velocity by projecting the orbital velocity of the gas,
\begin{equation}
v_{\phi, \textrm{gas}} \approx v_K + \frac{1}{2}\frac{c_s^2}{v_K}\frac{\partial \ln P}{\partial \ln r},
\end{equation}
in the $x$ and $y$ plane. In the above equation $v_K = \sqrt{GM_*/r}$ is the Keplerian orbital velocity, and $\partial \ln P / \partial \ln r$ is the radial gas pressure gradient. The radial gas pressure gradient is calculated as $\partial \ln P / \partial \ln r = \partial \ln \Sigma / \partial \ln r + \partial \ln T / \partial \ln r - \partial \ln H / \partial \ln r$, where $\partial \ln \Sigma / \partial \ln r = 15/14$. 


\subsection{Mass ablation}
If a planetesimal becomes sufficiently heated, then the material at its surface can undergo phase transitions, resulting in mass loss. Here it is assumed that the mass loss occurs due to solid material transitioning to gas phase by frictional heating and irradiation from the surrounding gas, a process called ablation.
In order to estimate the mass ablation rate from a planetesimal surface, we first need to know its composition. We assume that all planetesimals are non-differentiated, and that they consist of a mixture of silicate grains, carbon grains and volatile ices. In this paper we consider the three volatiles H$_2$O, CO$_2$ and CO, and the volatile content of a planetesimal is set by which volatile ices were present at its formation site. The total ablation rate from a planetesimal surface is then taken to be the sum of the ablation rates for each present volatile ice, and since the silicate and carbon grains are well mixed with the ices, we assume them to be released along with the ices (i.e. we assume no crust formation, see Appendix \ref{Appendix: noAbl} and \ref{Appendix: crust formation} for a discussion of this assumption). For a planetesimal which forms in a region of the disc where H$_2$O and CO$_2$ are in solid form, but CO is in gas form, the total ablation rate will thus be calculated as
\begin{equation}
\dot{m}_{\textrm{abl}} = \dot{m}_{\textrm{abl},\textrm{H}_2\textrm{O}} + \dot{m}_{\textrm{abl},\textrm{CO}_2}.
\end{equation}

We follow \citet{RonnetJohansen2020} and use the following expression for the ablation rate of an element $X$
\begin{equation}\label{eq:ablRate}
\dot{m}_{\textrm{abl},X} = -4\pi R_{\textrm{pl}}^2 P_{\textrm{sat},X}(T_{\textrm{pl}})\sqrt{\frac{\mu_{X}}{2\pi R_g T_{\textrm{pl}}}},
\end{equation}
where $P_{\textrm{sat},X}$ is the saturated vapor pressure of element $X$, $T_{\textrm{pl}}$ is the surface temperature of the planetesimal, $\mu_X$ is the molecular weight of element $X$, and $R_g$ is the ideal gas constant (e.g., \citealt{DAngeloPodolak2015}). Expressions for the saturated vapor pressure as polynomials of the temperature are given in \citet{FraySchmitt2009} for all three volatile ices under consideration. These polynomial expressions are only accurate above a certain temperature, which varies depending on the element. We therefore introduce "floor values" to the saturated vapor pressure. A plot of $P_{\textrm{sat},X}$ versus $T_{\textrm{pl}}$ for all three volatile ices, with the floor values included, is presented in Figure \ref{fig:Psat_Tpl} of the Appendix.

The equilibrium surface temperature of the planetesimals is obtained using the following equation from \citet{RonnetJohansen2020}
\begin{equation}\label{eq:Tpl}
T_{\textrm{pl}}^4 = T^4 + \frac{c_D \rho v_{\textrm{rel}}^3}{32\sigma_{\textrm{sb}}} - \sum_X \frac{P_{\textrm{sat},X}(T_{\textrm{pl}})}{\sigma_{\textrm{sb}}}\sqrt{\frac{\mu_X}{8\pi R_g T_{\textrm{pl}}}}L_X,
\end{equation}
where $\sigma_{\textrm{sb}}$ is the Stefan-Boltzmann constant, and $L_X$ is the latent heat of vaporization of element $X$. We calculated the latent heat of vaporization using the Clausius-Clapeyron relation for low temperatures and pressures, and obtained the results: $L_{\textrm{H}_2\textrm{O}} = 2.8\times 10^6\, \textrm{J}\, \textrm{kg}^{-1}$, $L_{\textrm{CO}_2} = 6.1\times 10^5\, \textrm{J}\, \textrm{kg}^{-1}$, and $L_{\textrm{CO}} = 2.8\times 10^5\, \textrm{J}\, \textrm{kg}^{-1}$. 
The second term in equation \ref{eq:Tpl} is heating due to gas friction, and the third term is cooling due to the ablation of volatile ices. Each volatile ice contributes its own cooling term, meaning that the planetesimal temperature will vary depending on what volatile ices it consists of. Since the cooling depends itself on the surface temperature of the planetesimal, the equation has to be solved iteratively. This is done using a bisection method. A plot of the mass ablation rate as a function of the semimajor axis is presented in Figure \ref{fig:ablRate_a_SS} of the Appendix for all three volatile ices, and for different surface temperatures of the planetesimals. 




\subsection{Condensation temperature}
We assume an isothermal equation of state for the disc pressure of an element $X$
\begin{equation}\label{eq:Pd(X)}
P_{d,X} = c_{s,X}^2 \rho_X.
\end{equation}
The sound speed of $X$ is calculated using equation \ref{eq:cs}, and exchanging the mean molecular weight of the disc with the molecular weight of $X$. The density of $X$ in the disc midplane is obtained by multiplying the total density in the midplane (equation \ref{eq:rhog0}) with the mass fraction of $X$ with respect to the disc. To obtain the relevant mass fractions we use abundances from \citet{Oberg2011}, and for the calculation we assume that 25\% of the disc mass is in He, and the remaining 75\% is in H. This results in the following mass fractions: $1.2\times 10^{-3}$, $9.7\times 10^{-4}$, and $3.1\times 10^{-3}$ for H$_2$O, CO$_2$ and CO. The mass fractions for silicate grains and carbon grains are $3.6\times 10^{-3}$ and $5.3\times 10^{-4}$. 

The condensation temperature of element $X$ is found by equating equation \ref{eq:Pd(X)} with the saturated vapor pressure, and solving for the temperature.
Since the density of the disc depends on $\dot{M}_0$, $\alpha$ and $r_{\textrm{out}}$, and these are parameters which are varied in between the simulations, the condensation temperatures will not be the same in all simulations. This means that a planetesimal which is initiated at the same semimajor axis in two separate simulations can have different compositions.

\subsection{Collision timescale}
We do not consider planetesimal-planetesimal collisions in our simulations; however, we still want an estimate of how common they are. If they happen on a timescale that is longer than the actual simulated time, it is justified to neglect them. However, if the opposite is true, we need to consider how they would have affected our results.

\citet{JohansenBitsch2019} provide an expression for the collision timescale of planetesimals in the gravitationally unfocused case
\begin{equation}\label{eq:tcoll}
t_{\rm coll} = \frac{1}{n_{\rm pl} \sigma_{\rm pl} \delta v} \approx \frac{R_{\rm pl} \rho_{\bullet}}{\Sigma_{\rm pl} \Omega} = \frac{R_{\rm pl} \rho_{\bullet}}{2 \delta v \rho_{\rm pl}},
\end{equation}
where $n_{\rm pl}$ is the number density, $\sigma_{\rm pl}$ is the physical cross section, $\delta v$ is the relative speed, and $\rho_{\rm pl} = \Sigma_{\rm pl} / (2H_{\rm pl})$ is the volume density of planetesimals. The relative speed between the planetesimals is approximated as
\begin{equation}\label{eq:deltav}
\delta v = \left(\frac{5}{4}e^2 + i^2\right)^{1/2}v_K
\end{equation}
\citep{LissauerStewart1993}. The scale height of the planetesimals is taken to be $H_{\rm pl} = i\times a$, where $a$ is the semimajor axis. 

\section{Numerical setup}\label{sect: numerical methods}
\begin{table*}
\caption{List of parameters for the simulations in the parameter study. Here $M_{\textrm{p}}$ denotes planetary mass, and "Comp." refers to the volatile composition of the planetesimals initiated at the gap edge of the specified planet.}
\label{table:paramStudy}
\centering
\begin{tabular}{lllllllll}
\hline\hline
\multicolumn{7}{l}{\textbf{Solar System}}                                                                                                                                  &           &             \\ \hline
Run                  & \multicolumn{2}{l}{\underline{Jupiter}}                   & \multicolumn{2}{l}{\underline{Saturn}}                     & $\alpha$                      & $\dot{M}_0$        &           &             \\
                     & $M_{\textrm{p}}$                          & Comp.          & $M_{\textrm{p}}$                           & Comp.          &                               &$(M_{\odot}\, \textrm{yr}^{-1})$             \\ \hline
\textbf{\#1} Nominal & $90\, \textrm{M}_{\oplus}$   & H$_2$O, CO$_2$ & $30\, \textrm{M}_{\oplus}$    & H$_2$O, CO$_2$ & $10^{-2}$                     & $10^{-7}$          &           &             \\
\textbf{\#2} Nominal (-CO$_2$)         & 90                           & H$_2$O         & 30                            & H$_2$O         & $10^{-2}$                     & $10^{-7}$          &           &             \\
\textbf{\#3} $M_{\textrm{p}}\downarrow$         & 30                           & H$_2$O, CO$_2$ & 10                            & H$_2$O, CO$_2$ & $10^{-2}$                     & $10^{-7}$          &           &             \\
\textbf{\#4} $M_{\textrm{p}}\uparrow$         & 150                          & H$_2$O, CO$_2$ & 50                            & H$_2$O, CO$_2$ & $10^{-2}$                     & $10^{-7}$          &           &             \\
\textbf{\#5} $\alpha\downarrow$         & 90                           & H$_2$O, CO$_2$ & 30                            & H$_2$O, CO$_2$ & $10^{-3}$                     & $10^{-7}$          &           &             \\
\textbf{\#6} $\dot{M}_0\downarrow$         & 90                           & H$_2$O & 30                            & H$_2$, CO$_2$         & $10^{-2}$                     & $10^{-8}$          &           &             \\ \hline
\multicolumn{9}{l}{\textbf{HL Tau}}                                                                                                                                                                  \\ \hline
Run                  & \multicolumn{2}{l}{\underline{Planet 1}}                  & \multicolumn{2}{l}{\underline{Planet 2}}                   & \multicolumn{2}{l}{\underline{Planet 3}}                       & $\alpha$  & $\dot{M}_0$ \\
\textbf{}            & $M_{\textrm{p}}$                          & Comp.          & $M_{\textrm{p}}$                           & Comp.          & $M_{\textrm{p}}$                           & Comp.              &           &$(M_{\odot}\, \textrm{yr}^{-1})$             \\ \hline
\textbf{\#7} Nominal & $59.6\, \textrm{M}_{\oplus}$ & H$_2$O, CO$_2$ & $141.2\, \textrm{M}_{\oplus}$ & H$_2$O, CO$_2$ & $313.7\, \textrm{M}_{\oplus}$ & H$_2$O, CO$_2$     & $10^{-2}$ & $10^{-7}$   \\
\textbf{\#8} $\alpha\downarrow$         & -//-                         & H$_2$O, CO$_2$ & -//-                          & H$_2$O, CO$_2$ & -//-                          & H$_2$O, CO$_2$, CO & $10^{-3}$ & $10^{-7}$   \\
\textbf{\#9} $\alpha\downarrow$ (-CO)         & -//-                         & H$_2$O, CO$_2$ & -//-                          & H$_2$O, CO$_2$ & -//-                          & H$_2$O, CO$_2$     & $10^{-3}$ & $10^{-7}$   \\
\textbf{\#10} $\alpha\downarrow$ (-CO, -CO$_2$)        & -//-                         & H$_2$O         & -//-                          & H$_2$O         & -//-                          & H$_2$O             & $10^{-3}$ & $10^{-7}$   \\
\textbf{\#11} $\dot{M}_0\downarrow$        & -//-                         & H$_2$O, CO$_2$ & -//-                          & H$_2$O, CO$_2$ & -//-                          & H$_2$O, CO$_2$     & $10^{-2}$ & $10^{-8}$   \\ 
\hline\hline
\end{tabular}
\end{table*}

We used the N-body code REBOUND to perform our simulations, and modified it to take into account the effect of gas drag and mass ablation \citep{ReinLiu2012}. The simulations were executed using the hybrid symplectic integrator MERCURIUS, and the timestep was set to be one twentieth of the innermost planet's dynamical timescale \citep{Rein2019}. Additional simulations with smaller timesteps were performed as well, in order to check that the outcome was not affected. We added the planets and the central star as active particles, and the planetesimals as semi-active particles with a mass. Semi-active particles only interact gravitationally with active particles. Collisions between active particles and semi-active particles are detected and recorded using a direct search method, and result in perfect merging. If a particle were to leave the simulation domain, it is recorded and removed from the simulation.  

We considered two planetary systems that would be representative of the young Solar and HL Tau systems. In the Solar System simulations the protoplanetary disc stretches from $0.1-100\, \textrm{au}$ with $R_{\textrm{out}}=20\, \textrm{au}$, and is modeled using a linear grid with 1000 grid cells. The location of a particle on this grid is found using a binary search algorithm. The simulation box is centered on the sun, and stretches $100\, \textrm{au}$ in $x$ and $y$-direction, and $20\, \textrm{au}$ in $z$-direction. In the HL Tau simulations we instead use a disc with 2000 grid cells, that stretches from $0.5-200\, \textrm{au}$ and has $R_{\textrm{out}}=100\, \textrm{au}$. The corresponding simulation box is $500\, \textrm{au}$ in $x$ and $y$-direction, and $100\, \textrm{au}$ in $z$-direction.

We use two planets in the Solar System simulations (Jupiter and Saturn), and three planets in the HL Tau simulations (placed at the locations of the major gaps in the disc). We do not consider planet growth or migration. Jupiter and Saturn are initiated with their current eccentricity and inclination, and we use their current bulk density to calculate the planetary radius, given the masses in Table \ref{table:paramStudy}. The three planets in HL Tau are initiated with close to zero eccentricity and inclination, and their radius is calculated using the masses in Table \ref{table:paramStudy} and assuming a constant density of $1000\, \textrm{kg}\,\textrm{m}^{-3}$. We use a central star of solar mass and solar luminosity in the HL Tau simulations.

The planetesimals are initiated uniformly between $1-2$ Hill radii away from the planets, in the direction away from the central star. This is roughly the region in which planetesimals form in Paper 1, and additional simulations show that small changes to this formation location does not affect the results. A study on how the simulation outcomes are affected by larger changes to the planetesimal formation location is presented in Appendix \ref{Appendix: gapwidth}. Generally, as long as the planetesimals do not form further away than about 5 Hill radii from the planets, the results does not change significantly. We initiate 50 planetesimals beyond each planetary gap, meaning that each individual Solar System simulation harbors 100 planetesimals, and each individual HL Tau simulation harbors 150. In order to provide better statistics we perform 10 simulations for each set of parameter that we study, so that the total number of planetesimals per parameter set amounts to 1000 and 1500 for the Solar System and HL Tau system simulations, respectively. 

The planetesimals are given an initial radius of $100\, \textrm{km}$, consistent with constraints from Solar System observations (\citealt{Bottke2005,Morbidelli2009}) and streaming instability simulations (e.g. \citealt{Johansen2015}), and have a constant solid density of $1000\, \textrm{kg}\, \textrm{m}^{-3}$. We add a property to the planetesimals that is the temperature of the disc at their formation location. This temperature, in relation to the condensation temperatures of the volatile ices, determines the composition of the planetesimals. The effect of gas drag is added as a velocity dependent force, and mass ablation is added as a post timestep modification. We keep track of how much mass is ablated from the planetesimals in each radial bin, and at what time during the simulation. If a planetesimal loses more than 99\% of its original mass, it is removed from the simulation and its remaining mass is considered to be ablated at the time and location where it was removed. The same procedure is applied if the amount of mass ablated in one timestep is larger than the total remaining mass of the planetesimal. 

We performed a parameter study in order to explore how the fate of the planetesimals formed at planetary gap edges is affected by: (1) their composition; (2) the mass of the planets; and (3) the density of the disc (controlled by the disc parameters $\alpha$ and $\dot{M}_0$). The parameter values used in the different simulations can be found in Table \ref{table:paramStudy}. 
In Simulation \#1, which we will hereafter refer to as the nominal Solar System simulation, the mass of Jupiter and Saturn was set to $90$ and $30\, M_{\oplus}$, $\alpha=10^{-2}$, and $\dot{M}_0=10^{-7}\, M_{\odot} \textrm{yr}^{-1}$. In such a disc the CO$_2$ iceline is located interior to the orbit of Jupiter, and the CO iceline is located well beyond the orbit of Saturn, meaning that all planetesimals in this simulation contain H$_2$O and CO$_2$ ice. The nominal HL Tau simulation is labeled \#7 in Table \ref{table:paramStudy}. In this simulation the planetary masses where set to equal the pebble isolation mass, calculated using an analytical fitting formula from \citet{Bitsch2018}. As in the nominal Solar System simulation, these planetesimals contain H$_2$O and CO$_2$ ice. 

\section{Simulations of the Solar System}\label{sect: solar system}
In the Solar System simulations we include two planets, Jupiter and Saturn, which are placed at their current semimajor axes. The mass ratio between Jupiter and Saturn is 3:1 in all simulations, which is roughly the current value, but the actual masses are varied.
Results from the nominal model are presented in Section \ref{subsect: SS nom}, and discussed in details. We examine how varying parameters affect the results in Section \ref{subsect: SS param}.

\subsection{Nominal model}\label{subsect: SS nom}
In the nominal model the mass of Jupiter is set to be $90\, M_{\oplus}$, and the mass of Saturn is set to be $30\, M_{\oplus}$. All parameters used in the nominal model, also called Simulation \#1, can be found in Table \ref{table:paramStudy}. The planetesimals are initiated in a narrow region just beyond the orbits of Jupiter and Saturn, as suggested by the streaming instability simulations in Paper 1. At these distances from the Sun, and with the disc parameters stated in Table \ref{table:paramStudy}, H$_2$O and CO$_2$ are in solid form while CO is in gas phase. We therefore consider these planetesimals to have a volatile content of H$_2$O and CO$_2$, and consider the ablation of both these molecules.

\subsubsection{Dynamical evolution}\label{subsubsect: SS nom dyn ev}

\begin{figure*}
\centering
\resizebox{0.7\hsize}{!}
    {\includegraphics{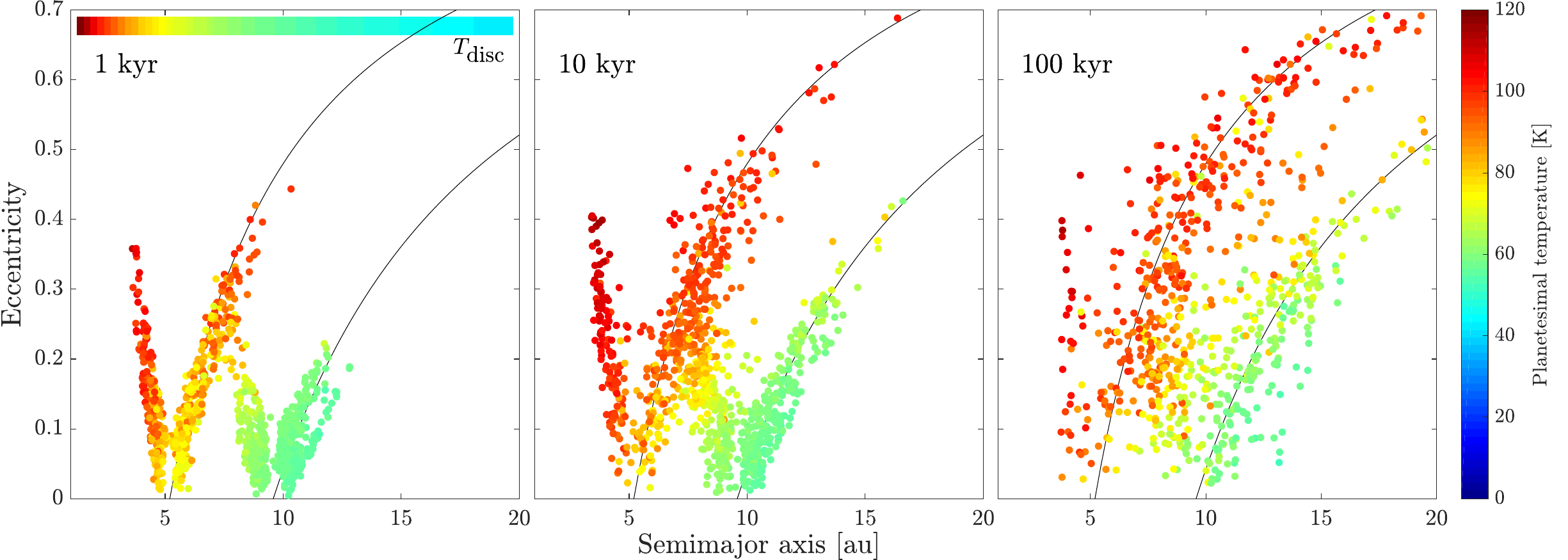}}
\caption{Eccentricity, semimajor axis and surface temperature evolution for the 1000 planetesimals in the nominal Solar System simulation. The presented orbital parameters have been averaged over $100\, \textrm{yr}$, and the surface temperatures are the maximum values during the same time period, which roughly corresponds to the surface temperatures at perihelion. The temperature of the surrounding gas ($T_{\textrm{disc}}$) is shown as a colorbar in the left panel. The solid black lines mark the orbits with a perihelion corresponding to Jupiter’s and Saturn’s location. The planetesimals which cluster around these lines are (at least momentarily) members of Jupiter's or Saturn's scattered disc. Planetesimals which are scattered interior of Jupiter's orbit obtain high surface temperatures at perihelion and experience efficient ablation.}
    \label{fig:e_a_T_nominal_SS}
\end{figure*}

\begin{figure*}
\centering
\resizebox{0.7\hsize}{!}
    {\includegraphics{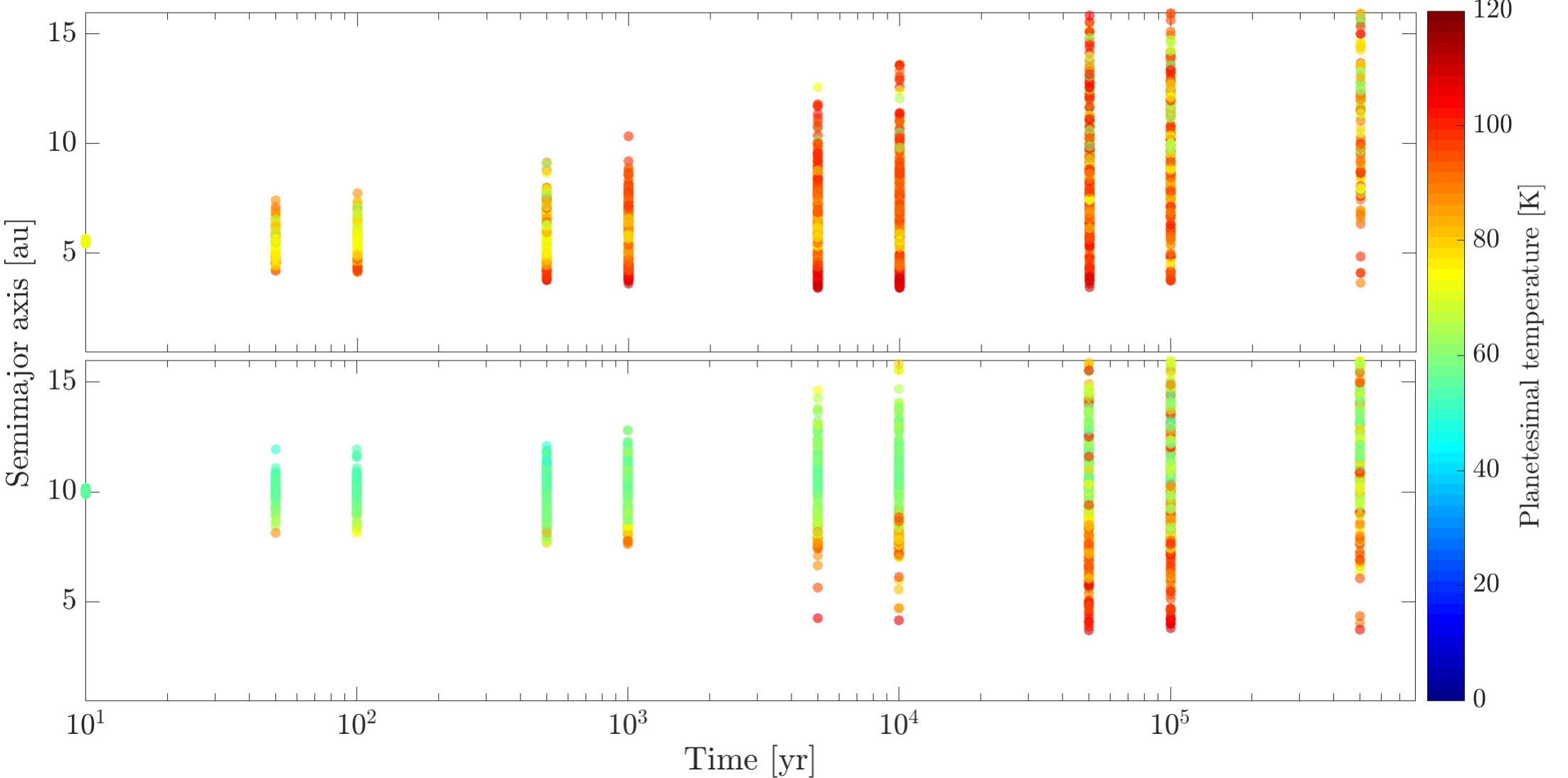}}
\caption{Semimajor axis and surface temperature evolution for the 1000 planetesimals in the nominal Solar System simulation (same data as in Figure \ref{fig:e_a_T_nominal_SS}). For simulation times longer than $1\, \textrm{kyr}$, we average the semimajor axes over $100\, \textrm{yr}$ and show the maximum surface temperatures during the same time period; while for shorter simulations times, we show non-averaged values. The upper panel contains planetesimals formed at the gap edge of Jupiter, and the lower panel contains planetesimals formed at the gap edge of Saturn. The close proximity to the giant planets result in continuous scatterings, which causes the fast semimajor axis diffusion that is displayed in the plot.}
    \label{fig:a_t_T_nominal_SS}
\end{figure*}

In this section we address the question of the dynamical redistribution of the planetesimals formed at planetary gap edges. In Figure \ref{fig:e_a_T_nominal_SS} and \ref{fig:a_t_T_nominal_SS} we show the eccentricity and semimajor axis evolution for the 1000 planetesimals in the nominal simulation. The planetesimals are initiated with an eccentricity close to zero, but from Figure \ref{fig:e_a_T_nominal_SS} we see that the close proximity to the planets leads to rapid scattering, resulting in eccentricities as high as 0.4 already after $1\, \textrm{kyr}$ of evolution. Most planetesimals that are not immediately scattered interior to the orbit of Jupiter end up in the scattered disc of Jupiter or Saturn. The eccentricities within these scattered discs increases with time, as the planetesimals are continuously scattered at perihelion. Such large eccentricities lead to large velocities at perihelion ($v_{\textrm{peri}}$) relative to the gas,
\begin{equation}
v_{\textrm{peri}} = v_K \sqrt{\frac{1+e}{1-e}},
\end{equation}
which in turn affects the thermodynamic evolution of the planetesimals. The velocity at perihelion is higher within Jupiter's scattered disc than it is within Saturn's for the same orbital eccentricity, since the Keplerian velocity decreases with increasing semimajor axis. 

In Figure \ref{fig:a_t_T_nominal_SS} we have separated the planetesimals forming at the gap edge of Jupiter and Saturn into different panels, making the diffusion of the semimajor axes clearly visible. Strong scatterings causes the semimajor axes to diffuse over several au in less than $100\, \textrm{yr}$. After a few thousand years planetesimals from both gap edges are spread out across the giant planet region, and the ones closest to the star have semimajor axes of $\sim\!\! 3\, \textrm{au}$. A small number of planetesimals, $13/1000$ in this particular simulation, suffer a collision with either Jupiter or Saturn. Most of these collisions are between Jupiter and planetesimals formed at Jupiter's gap edge. Concerning the planetesimals that obtain very large eccentricities and semimajor axes, many of them are eventually scattered outside the simulation domain, and are considered to be ejected from the system. In this simulation $\sim\!\! 15\%$ of all planetesimals are eventually ejected. 

In summary, planetesimals which form at the edges of planetary gaps do not remain at their initial birth location. The close proximity to the gap-opening planets results in strong scatterings, causing the planetesimals to spread out across the protoplanetary disc. Many planetesimals initially become members of the gap-opening planet's scattered disc, while some are scattered towards the inner disc region. The resulting high eccentricities lead to a high velocity relative to the gas, which has big implications for the mass evolution as will be seen in Sect. \ref{subsubsect: SS nom mass ev}. In Appendix \ref{Appendix: gapwidth} we show simulations where the planetesimals are formed further from the planet. Placing planetesimals further from the planet does not change our results qualitatively, but does lead to a delayed and slower scattering phase. 

\subsubsection{Mass evolution}\label{subsubsect: SS nom mass ev}

\begin{figure}
\centering
    {\includegraphics[width=\hsize]{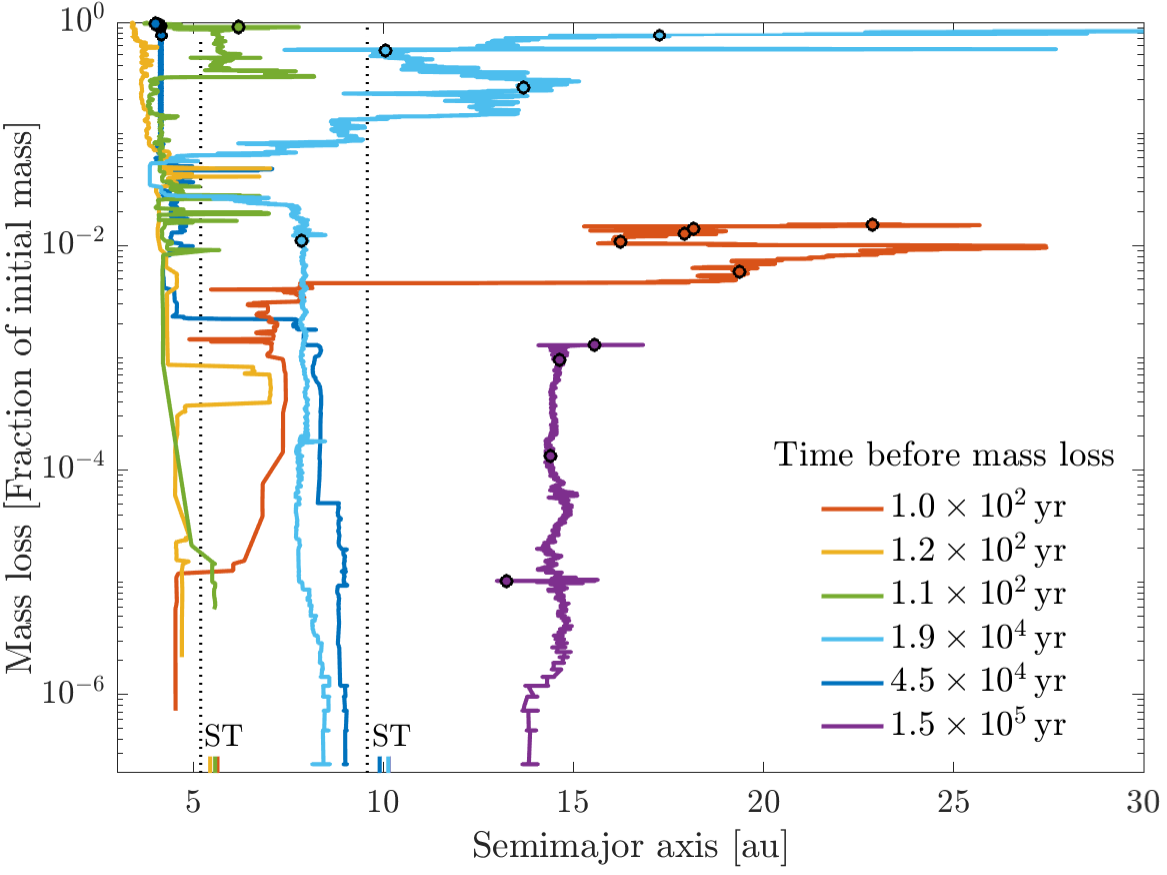}}
\caption{Plot of mass loss versus semimajor axis for 6 selected planetesimals from the nominal Solar System simulation. The mass loss is calculated as $1-M(t)/M(t=0)$. Filled circles mark $100\, \textrm{kyr}$ of evolution, and the formation location of the planetesimals (shown at the bottom of the plot) has been marked "ST". The time before the planetesimals first experience mass loss, that is the time before they appear on the plot, has been written in the legend. The first three legend entries are for planetesimals formed at the gap edge of Jupiter, and the following three are for planetesimals formed at the gap edge of Saturn. The dotted lines mark the semimajor axis of Jupiter and Saturn. Planetesimals which are scattered interior of Jupiter's orbit lose mass at a high rate, while those which are scattered exterior of Saturn's orbit experience little mass loss. }
	\label{fig:ml_a_6plan_nominal_SS}
\end{figure}

\begin{figure*}
\centering
    {\includegraphics[width=1.7\columnwidth]{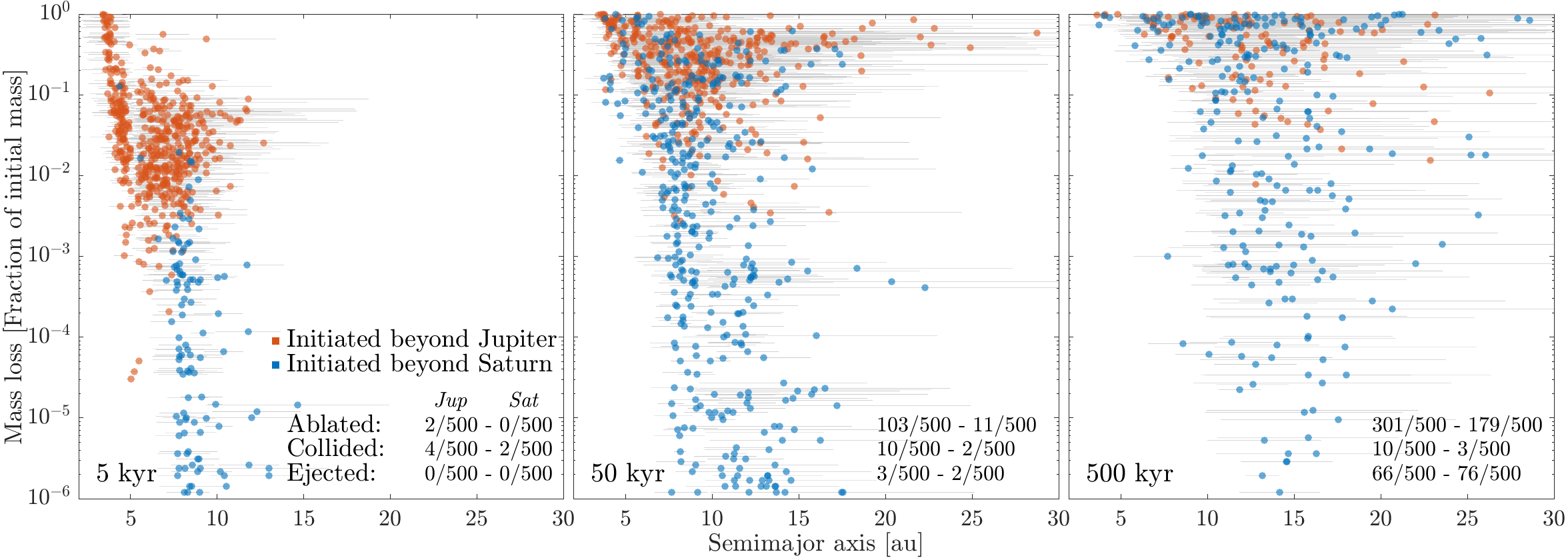}}
\caption{Mass loss versus semimajor axis evolution for the 1000 planetesimals in the nominal Solar System simulation (same data as in Figure \ref{fig:e_a_T_nominal_SS} and \ref{fig:a_t_T_nominal_SS}). The thin grey lines mark the perihelion and aphelion of the planetesimal orbits. The number of planetesimals formed at Jupiter's (red dots) respective Saturn's (blue dots) gap edge which do not appear on the plot, because they have been either: completely ablated; collided with a planet; or ejected beyond the simulations domain, is written in each panel. Planetesimals formed at the gap edge of Jupiter generally experience more ablation than planetesimals formed at the gap edge of Saturn. About 50\% of all planetesimals have become completely ablated after $500\, \textrm{kyr}$. }
    \label{fig:ml_a_nominal_SS}
\end{figure*}

In this section we look into the mass evolution of planetesimals; that is, the loss of mass due to ablation. The ablation rate depends strongly on the planetesimal surface temperature, which increases towards the central star as the disc temperature and gas density becomes higher. The surface temperature is also highly dependent on the velocity relative to the gas, which increases with increasing orbital eccentricity as described in Sect. \ref{subsubsect: SS nom dyn ev}. Based on this, the highest surface temperature, and thus ablation rates, should be obtained at the perihelion passage of eccentric orbits close to the star. This can be seen in Figure \ref{fig:e_a_T_nominal_SS} and \ref{fig:a_t_T_nominal_SS}, where we also show the surface temperature evolution of the planetesimals at perihelion. The planetesimals that are scattered interior to Jupiter's orbit obtain surface temperatures around $100\, \textrm{K}$ at perihelion. At these temperatures CO$_2$ ablation quickly sublimates the planetesimal, which is why there are no planetesimals in the innermost part of the disc. Note here that the ablation process itself severely decreases the surface temperatures through cooling due to the latent heat of vaporization. A similar simulation without taking into account ablation would result in many times higher surface temperatures (see Appendix \ref{Appendix: noAbl}). Note also that a planetesimal with the same surface temperature and orbital parameters in a gas-free disc would not experience any ablation due to the lack of frictional heating.

In Figure \ref{fig:ml_a_6plan_nominal_SS} we track the mass loss of 6 selected planetesimals from the nominal simulation, as they are scattered around by the planets. The time evolution of the semimajor axes, perihelia, aphelia and eccentricity for the same planetesimals can be found in Figure \ref{fig:a_e_t_6plan_nom_SS} of the Appendix. 
The planetesimals that form at Jupiter's gap edge (the first three legend entries), begin to experience mass loss already after $100\, \textrm{yr}$ of evolution. The "red" planetesimal is a member of Jupiter's scattered disc for a few thousand years, until it is kicked towards Saturn's scattered disc, where it remains until the end of the simulation. Since its orbit never enters the inner disc region, ablation is slow and the planetesimal only loses about 2\% of its mass. The "yellow" planetesimal obtains multiple kicks by Jupiter during the first few thousand years, after which it ends up interior to Jupiter's orbit with a perihelion of just above $2\, \textrm{au}$, and becomes completely ablated within $10\, \textrm{kyr}$. The "green" planetesimal is a member of Jupiter's scattered disc for $100\, \textrm{kyr}$, with a relatively small eccentricity, until a strong planetary encounter leaves it on an orbit interior to Jupiter, where it very quickly becomes ablated.

The planetesimals formed at Saturn's gap edge (the last three legend entries), begin to lose mass much later. The "light-blue" planetesimal sits on a low-eccentric orbit in between Jupiter and Saturn for $150\, \textrm{kyr}$, after which it obtains a strong kick and becomes a high-eccentric member of Jupiter's scattered disc. Its eccentricity continues to increase until it leaves the simulation domain, having lost 85\% of its mass. The "dark-blue" planetesimal is scattered onto a $4\, \textrm{au}$ orbit by Saturn, where it slowly becomes circularized. Once circularized, the relative velocity between the planetesimal and the gas turns to zero, and it remains in the same orbit until the end of the simulation. Finally, the "purple" planetesimal never enters the region interior to Saturn, but remains on a low-eccentric orbit in the outer part of the disc for the entire simulation.

The mass loss tracks presented in Figure \ref{fig:ml_a_6plan_nominal_SS} are examples of what can happpen to a planetesimal in the simulation. In Figure \ref{fig:ml_a_nominal_SS} we present the mass loss and semimajor axis evolution for all planetesimals in the nominal simulation. The planetesimals formed at Jupiter's gap edge start to lose mass much earlier than those formed at Saturn's gap edge, and they generally lose mass at a faster rate. This is very much expected, mostly because the planetesimals form in a warmer part of the disc; but also because Saturn is of lower mass than Jupiter, and therefore does not scatter the planetesimals as strongly. Towards the end of the simulation, $\sim\!\! 50\%$ of all planetesimals have become completely ablated. In Appendix \ref{Appendix: gapwidth} we show that placing the planetesimals further from the planet results in a delayed and slower ablation phase.

From the plots that have been presented it is evident that mass loss due to ablation plays a major role in the evolution of planetesimals formed at planetary gap edges, at least for the parameters used in the nominal model. In Section \ref{subsect: SS param} we will investigate exactly how much mass is lost and where and compare that to simulations with varying planetary masses and disc parameters.

\subsubsection{Planetesimal-planetesimal collisions}

\begin{figure}
\centering
    {\includegraphics[width=.8\hsize]{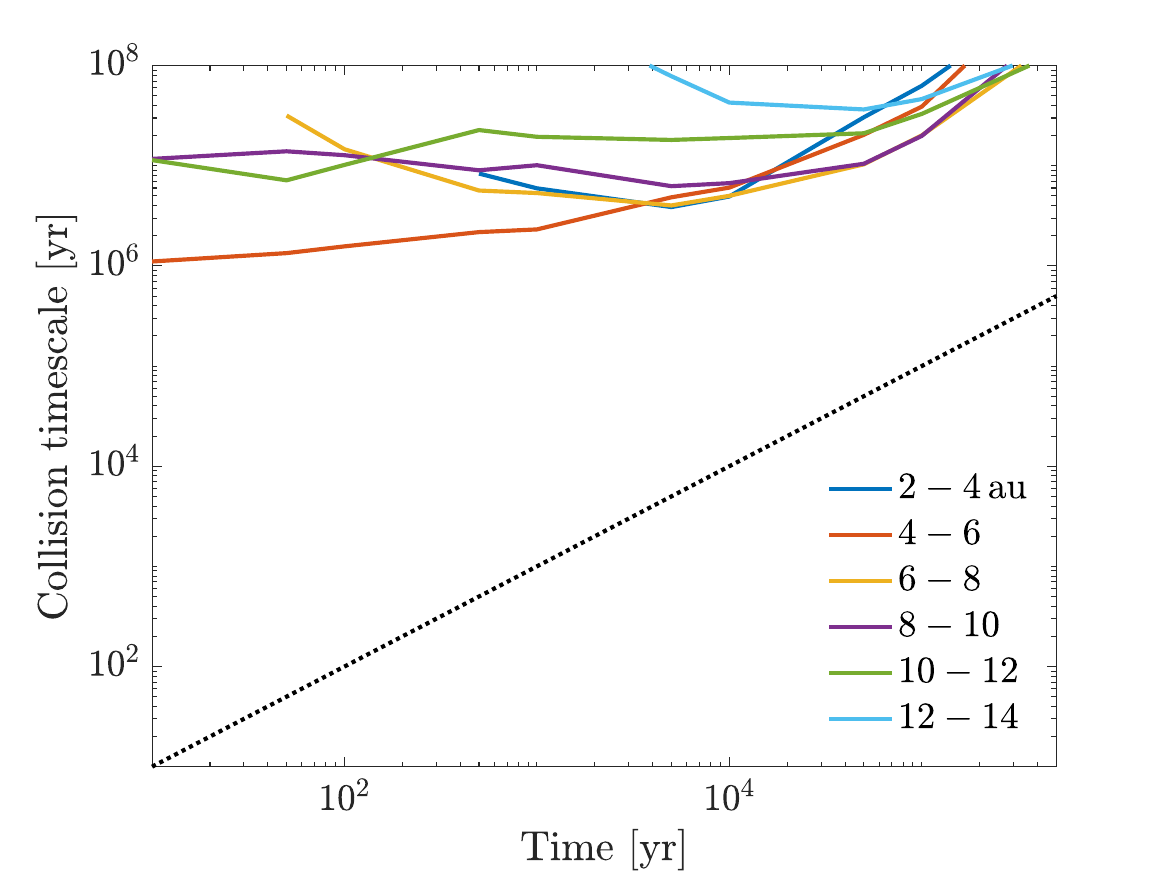}}
\caption{Time evolution of the timescale for planetesimal-planetesimal collisions for the nominal Solar System simulation. The collision timescale was calculated using equation \ref{eq:tcoll} and assuming an initial planetesimal mass of $16\, \textrm{M}_{\oplus}$ per gap edge. The dotted line marks where the collision timescale equals the time of the simulation. Since the resulting collision timescale for one planetesimal is several orders of magnitude larger than the actual simulated time, planetesimal-planetesimal collisions can be safely ignored in our simulations and will not constitute a significant path to replenishing the dust component in the disc. }
    \label{fig:collTime_nominal_SS}
\end{figure}

In this section we use simple calculations to estimate how common planetesimal-planetesimal collisions are. If they occur on the same timescale as dynamical scattering then they could affect our results. Furthermore, if these collisions are strong enough to disrupt the planetesimals that are involved, it would constitute another mechanism to replenish dust and pebbles in the disc.

We split the disc into multiple semimajor axis rings and find the planetesimals which are located within each ring. We then use the average eccentricity and inclination of those planetesimals to get an estimate of the velocity dispersion (equation \ref{eq:deltav}) and the scale height. To calculate the collision timescale, we further need to know the volume density of planetesimals within each ring. If all solids between the semimajor axis of Jupiter and Saturn is converted into planetesimals at Jupiter's gap edge, then there would be $16\, \textrm{M}_{\oplus}$ of planetesimals at that location, using the disc parameters of the nominal model and a dust-to-gas ratio of 1\%. We assume that the amount of planetesimals forming at Saturn's gap edge is the same. The total planetesimal mass in a ring is then simply taken to be the sum of the mass of all planetesimals in that ring.

When using a ring-width of $2\, \textrm{au}$, this resulted in collision timescales above $1\, \textrm{Myr}$, at all locations in the disc and at all times during the simulation (see Figure \ref{fig:collTime_nominal_SS}). Decreasing the ring-width to $1\, \textrm{au}$ lead to slightly smaller collision timescales. These results indicate that an individual planetesimal suffer a very low risk of colliding with another planetesimal. Since in reality there are millions of planetesimals forming at the gap edges, collisions will still be occurring, but not often enough to affect any of our results.

\subsection{Parameter study and mass loss profiles}\label{subsect: SS param}

\begin{figure*}
\centering
    {\includegraphics[width=\columnwidth]{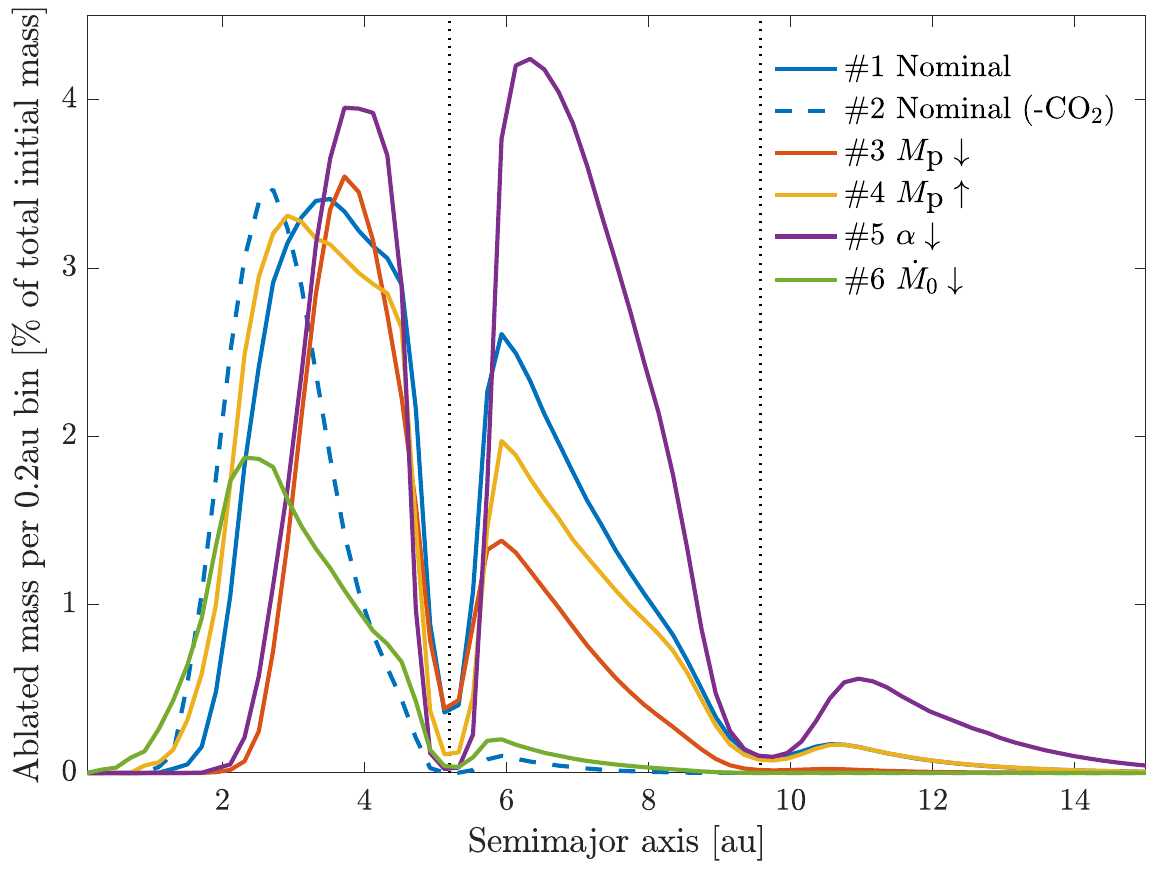}}
    {\includegraphics[width=\columnwidth]{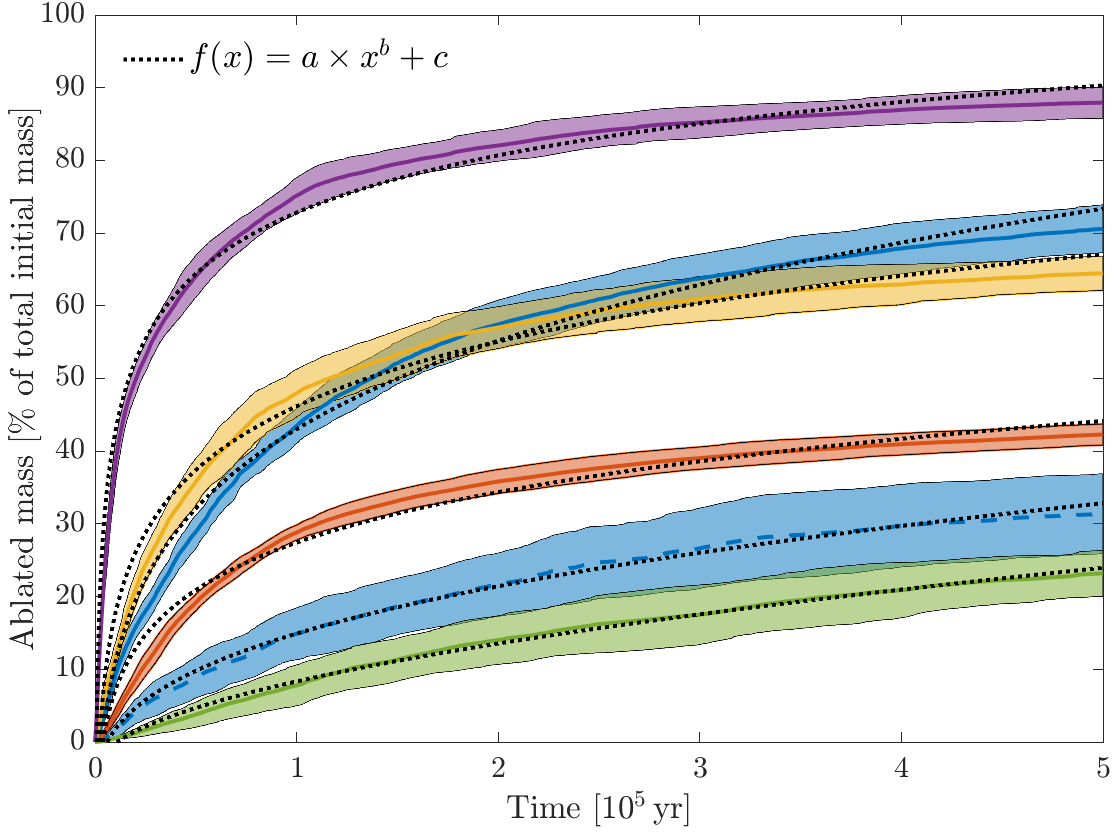}}
\caption{Left plot: distribution of ablated mass across the disc after $500\, \textrm{kyr}$ for all Solar System simulations. The values on the $y$-axis represent the amount of mass that has been ablated in a $0.2\, \textrm{au}$ semimajor axis bin. The dotted lines mark the semimajor axes of Jupiter and Saturn. Most mass loss occur in the region just interior to and exterior to Jupiter's orbit. Efficient ablation interior of Jupiter's orbit results in that the planetesimals never reach the innermost disc region. Right plot: the total amount of mass that has been ablated as a function of time, for the same data as in the left plot. The colored lines show the average over the 10 simulations, and the colored region shows the one standard deviation away from this value. The dotted black lines show the best fit to the curve $f(x)$; and the parameters to the fits can be found in Table \ref{table:param SS}. The mass ablation rate is highly dependent on the gas surface density, where a high surface density results in much ablation (\#5), and a low surface density results in little ablation (\#6). }
	\label{fig:abl_SS}
\end{figure*}

\begin{table}
\centering
\caption{Parameters to the fitted curve $f(x)=a\times x^b+c$ which is shown in the right panel of Figure \ref{fig:abl_SS}. }
\label{table:param SS}
\begin{tabular}{llll}
\hline\hline
Run & $a$    & $b$     & $c$     \\ \hline
\textbf{\#1} Nominal   & 108.1  & 0.1537  & -65.02  \\
\textbf{\#2} Nominal (-CO$_2$)   & 23.9   & 0.3476  & -9.02   \\
\textbf{\#3} $M_{\textrm{p}}\downarrow$   & 117.7  & 0.0826  & -90.24  \\
\textbf{\#4} $M_{\textrm{p}}\uparrow$   & 572.5  & 0.0223  & -526.30 \\
\textbf{\#5} $\alpha\downarrow$   & -120.4 & -0.0976 & 193.20  \\
\textbf{\#6} $\dot{M}_0\downarrow$   & 12.29  & 0.5099  & -3.96   \\ \hline
\end{tabular}
\end{table}

\begin{figure*}
\centering
\resizebox{\hsize}{!}
{\includegraphics{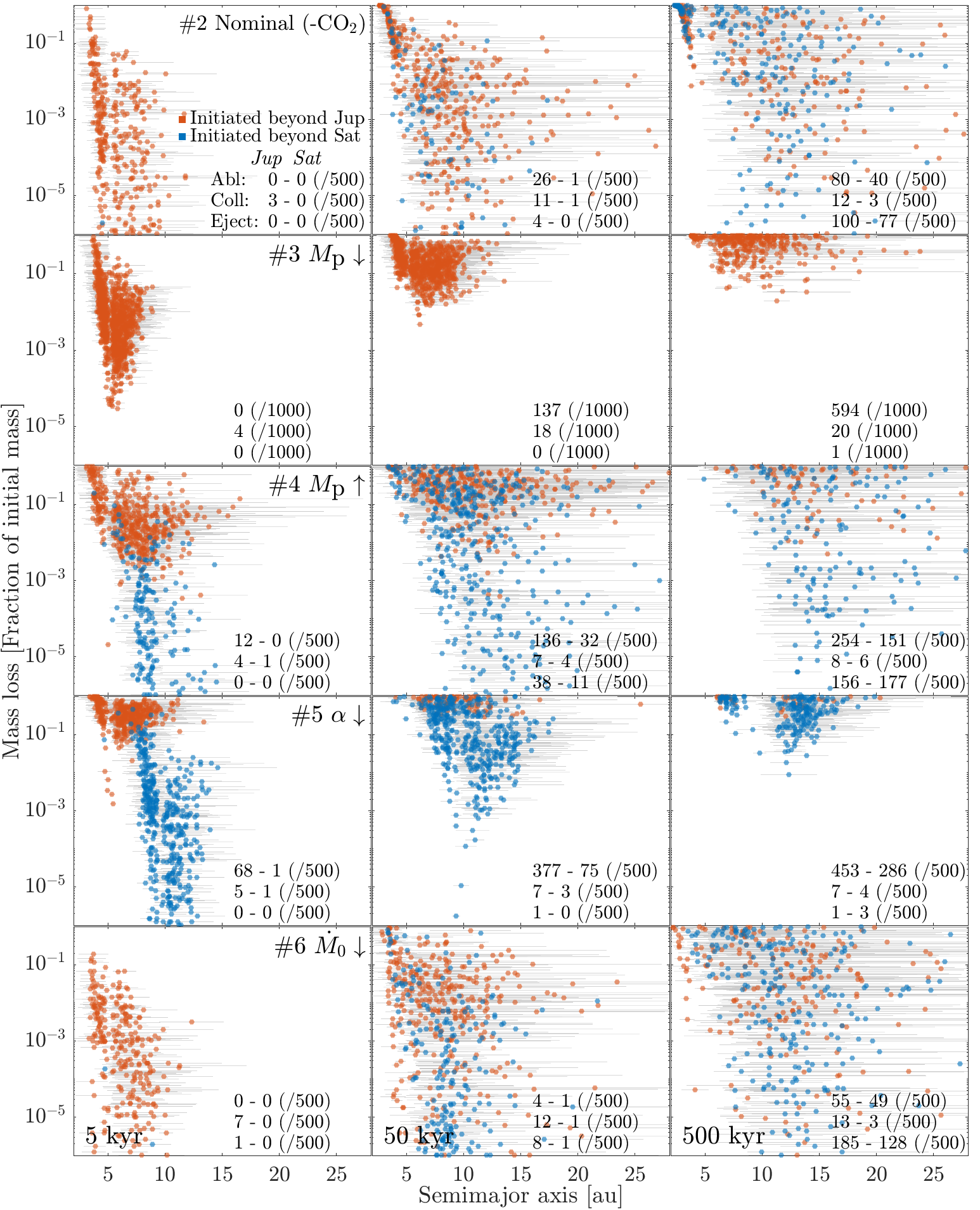}}
\caption{Mass loss versus semimajor axis evolution for all Solar System simulations. The masses of the planets affect how far the planetesimals are scattered, with more planetesimals leaving the simulation domain when the planetary mass is increased (\#4). A large surface density (\#5) generally results in efficient ablation and few scatterings beyond the simulation domain, since the planetesimals are ablated before they become scattered. The opposite is the case for a low surface density (\#6), and ablation also occurs closer to the star. }
    \label{fig:ml_a_allParam_SS}
\end{figure*}

In this section we vary certain disc and planet parameters in order to find out how the simulation results are affected. The parameters used in each simulation can be found in Table \ref{table:paramStudy}. In Figure \ref{fig:abl_SS} we show how the total amount of ablated mass varies across the disc (left panel), and as a function of time (right panel). The mass loss evolution for each simulation is shown in Figure \ref{fig:ml_a_allParam_SS}, where the number of planetesimals becoming completely ablated; scattered; or suffered planetary collisions can be read off and compared as well. The collision timescale for all simulations in the parameter study are very similar to those in the nominal simulation, meaning that planetesimal-planetesimal collisions can be safely ignored as a means to produce dust.

In the nominal simulation, 70\% of the initial planetesimal mass has been ablated at the end of the simulation. The majority of this mass is released either just interior of Jupiter's orbit, between $2-5\, \textrm{au}$, or just exterior of it, between $6-8\, \textrm{au}$. Mass loss exterior of Jupiter's orbit is possible because of the large orbital eccentricities, resulting in high surface temperatures. The planetesimals in the nominal simulation all form beyond the CO$_2$ iceline, and thus both H$_2$O and CO$_2$ ablation is considered. In Simulation \#2 we remove CO$_2$ ablation, and consider the volatile content of the planetesimals to be only H$_2$O. As a result, the amount of ablated mass at the end of the simulation drops from 70\% to 30\%. Furthermore, since the ablation rate of H$_2$O is lower than that of CO$_2$ for the same temperature (see Figure \ref{fig:ablRate_a_SS} in Appendix), the planetesimals need to be scattered further towards the Sun in order for efficient ablation to occur. This can be seen in the left panel of Figure \ref{fig:abl_SS}, where the peak of the ablation curve interior to Jupiter has been shifted closer to the Sun. Additionally, there is no longer any mass released beyond the orbit of Jupiter, telling us that all mass released in this region in the nominal simulation was because of the presence of CO$_2$.

In Simulation \#3 we decrease the planetary masses by a factor of 3, mimicking an earlier stage of the planet formation process. The mass of Jupiter is now equal to the pebble isolation mass, while the mass of Saturn is only 20\% of the pebble isolation mass. In Paper 1 the amount of planetesimals that form at the gap edges drop by 80\% when the mass is decreased from one pebble isolation mass to half a pebble isolation mass. The amount of planetesimals forming beyond Saturn in this simulation would thus have been very small, if any at all, and so we only include planetesimals at Jupiter's gap edge. The number of planetesimals in the simulation is kept the same as in all other simulations, but since they represent only half the mass, we divide the amount of ablation by a factor of 2 before presenting the results in Figure \ref{fig:abl_SS}. 

Lowering the planetary masses and removing the planetesimals at Saturn's gap edge result in about 40\% less mass ablation than in the nominal simulation. The removal of planetesimals at Saturn's gap edge results in fewer planetesimals far out in the disc. Lowering the planetary masses results in weaker planetary scatterings, which in turn has several effects: (1) only 1/1000 planetesimals are ejected beyond the simulation domain; (2) the orbits are not as excited, resulting in lower surface temperatures and slower ablation; (3) the planetesimals end up on orbits closer to their birth locations, and thus the mass loss is more concentrated towards Jupiter's location. 

In Simulation \#4, the planetary masses are increased, mimicking a later stage in the planet formation process. The stronger planetary scatterings initially result in higher mass loss rates, as the planetesimals orbits quickly become excited. However, the strong planetary scatterings also result in that many more planetesimals are ejected beyond the simulation domain, which leaves less planetesimals in the system to be ablated during later times. 

In the final two simulations of the parameter study, Simulation \#5 and \#6, the viscosity parameter and disc accretion rate are decreased by a factor of 10. Decreasing the viscosity parameter slightly changes the gap profiles, but mostly it results in a ten times larger surface density, thus mimicking an earlier stage of disc evolution. The ice lines are also shifted closer to the Sun, but in this case that does not change the composition of the planetesimals compared to the nominal model. On the other hand, decreasing the disc accretion rate results in a ten times lower surface density, causing the ice lines to move further out in the disc. As a result the CO$_2$ ice line ends up beyond the gap edge of Jupiter, and thus the planetesimals forming there do not contain any CO$_2$ ice. 

Looking at Figure \ref{fig:abl_SS}, we see that lowering the viscosity parameter results in more ablation, while lowering the disc accretion rate results in less ablation, just as expected. The higher surface density in Simulation \#5 results in higher surface temperatures, allowing for planetesimals to lose mass further out in the disc, and at a higher rate. This in a combination with larger gas friction results in that almost no planetesimals are ejected from the system. The opposite is the case in Simulation \#6, and since the planetesimals formed at Jupiter's gap edge do not contain CO$_2$ ice, there is little mass loss exterior of Jupiter's orbit, just like in Simulation \#2. 

The results from the parameter study show that the surface density of the disc is a key parameter in determining how much mass is ablated from the planetesimals. Since the surface density in discs decreases with time, ablation is expected to be much more efficient in young discs than in old ones. Planetesimals forming at the gap edge late during the disc lifetime might thus not suffer any significant mass loss due to ablation, but could for example be implanted into the asteroid belt or aid in the water delivery to Earth \citep{RaymondIzidoro2017}. The masses of the planets determine how excited the planetesimal orbits become and how far they are scattered within the disc. To some extent the planetary mass also affects how many planetesimals that form at the gap edge (see Paper 1). Finally, the composition of the planetesimals plays a major role in determining how efficient ablation is. This is set by the location of formation relative to the location of the major ice lines, which in a real disc shifts inwards with time as the temperature of the disc decreases.

\section{Simulations of HL Tau}\label{sect:HL Tau}
We include three planets in the HL Tau simulations and place them at the locations of the major gaps in the disc, that is at $11.8$, $32.3$ and $82\, \textrm{au}$ \citep{Kanagawa2016}. The mass of the planets is set to be $59.6$, $141.2$ and $313.7\, M_{\oplus}$, which is exactly equal to the pebble isolation mass using the parameters in Simulation \#7. For reference, this is the same set-up as in Simulation \#4 of Paper 1. Each simulation contains 150 planetesimals, with 50 formed at each planetary gap edge, and we run 10 simulations per parameter set. Results from the nominal simulation are presented in Section \ref{subsect: HL Tau nom}, and in Section \ref{subsect: HL Tau param} we study how these results change when we vary the planetesimal compositions and the disc parameters.

\subsection{Nominal model}\label{subsect: HL Tau nom}
The three planets in the HL Tau disc are located well beyond the CO$_2$ iceline, which in the nominal model (Simulation \#7) is at $4.7\, \textrm{au}$. The CO iceline sits much further out in the disc at $99.3\, \textrm{au}$, placing it just beyond the formation location of the outermost planetesimal. Since all planetesimals in the nominal model form in between the CO$_2$ and CO iceline, we thus consider them to be made up of H$_2$O and CO$_2$ ice, and consider the ablation of these two molecules. 

\begin{figure*}
\centering
\resizebox{0.7\hsize}{!}
    {\includegraphics{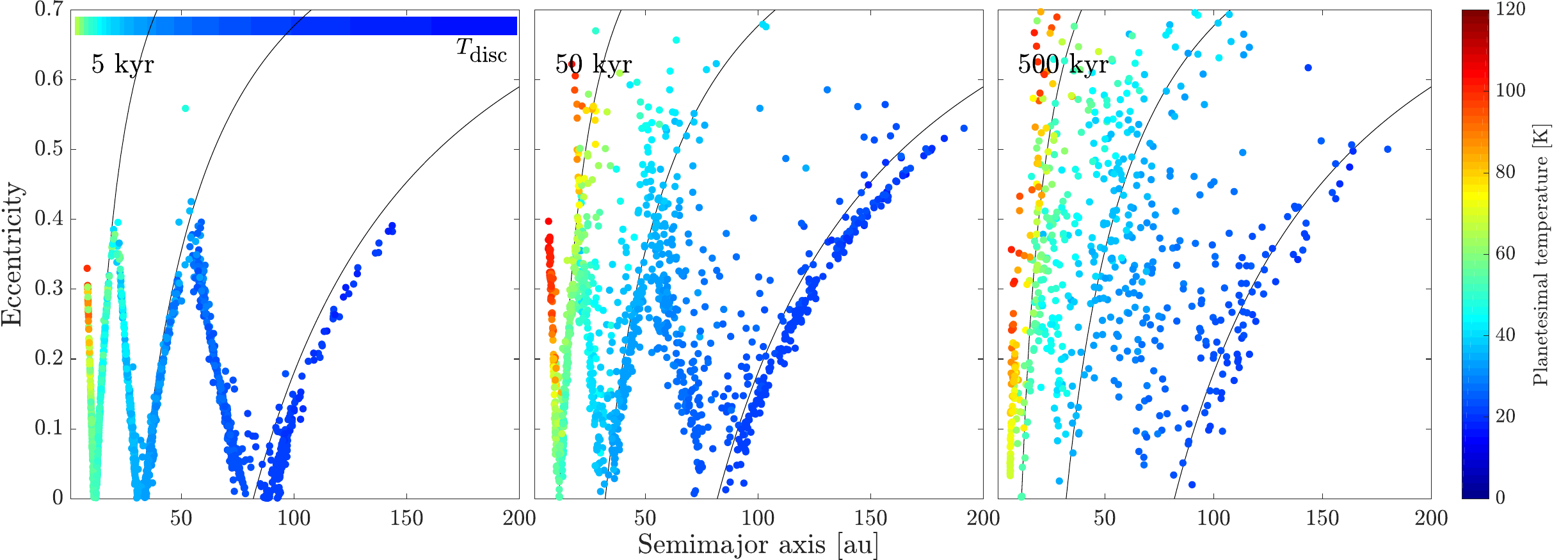}}
\caption{Eccentricity, semimajor axis and surface temperature evolution for the 1500 planetesimals in the nominal HL Tau simulation, produced in a similar manner as Figure \ref{fig:e_a_T_nominal_SS}. The solid black lines mark the perihelia of the planets. Many planetesimals end up in the scattered discs of the planets, where they obtain high eccentricities. Only planetesimals scattered to interior of the innermost planet's orbit obtain perihelion temperatures large enough for ablation to take place.}
    \label{fig:e_a_T_nominal_HLTau}
\end{figure*}

\begin{figure*}
\centering
\resizebox{0.8\hsize}{!}
    {\includegraphics{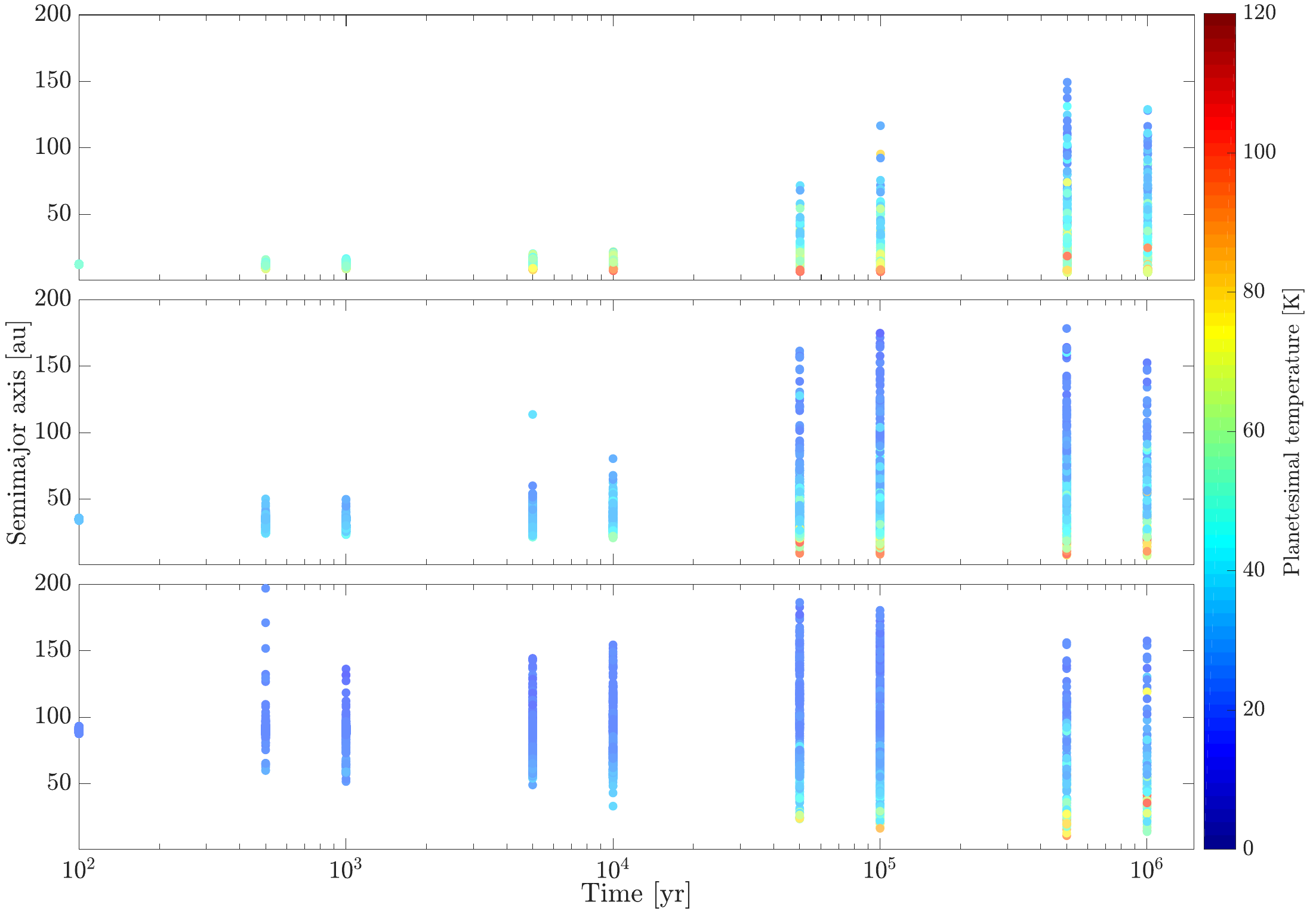}}
\caption{Semimajor axis and surface temperature evolution for the 1500 planetesimals in the nominal HL Tau simulation, produced in a similar manner as Figure \ref{fig:a_t_T_nominal_SS} (same data as in Figure \ref{fig:e_a_T_nominal_HLTau}). Planetesimals formed at the gap edge of the innermost planet is shown in the upper panel, and similar for the middle and outermost planet (middle respective bottom panels). Since the planetary masses are set to equal the pebble isolation mass, and the pebble isolation mass increases with semimajor axis, the planetesimals formed at the gap edge of the outermost planet experience stronger scatterings and diffuse faster than those formed at the innermost planet's gap edge. }
    \label{fig:a_t_T_nominal_HLTaul}
\end{figure*}

The dynamical evolution of the 1500 planetesimals from the nominal simulation is presented in Figure \ref{fig:e_a_T_nominal_HLTau} and \ref{fig:a_t_T_nominal_HLTaul}. Just like in the Solar System simulations, most planetesimals end up in the scattered discs of the planets, with eccentricities increasing over time. The two outermost planets are very massive and deliver strong kicks to the planetesimals formed in their vicinity, causing them to quickly spread out over the entire disc. Many of these planetesimals are eventually ejected beyond the simulation domain. The planetesimals which form at the gap edge of the innermost planet instead suffer weaker kicks, and tend to remain in the inner disc region.

\begin{figure}
\centering
    {\includegraphics[width=.7\hsize]{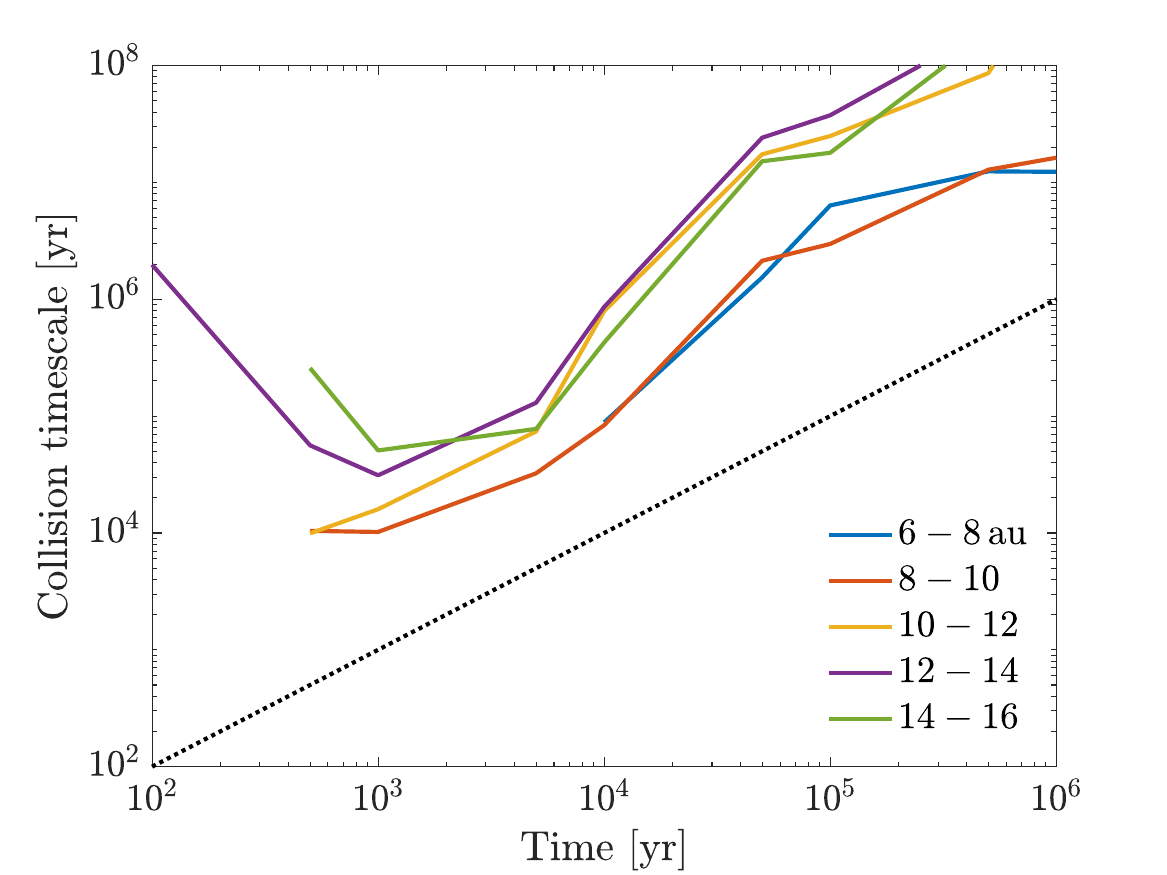}}
\caption{Time evolution of the timescale for planetesimal-planetesimal collisions for the nominal HL Tau simulation. The collision timescale was calculated using equation \ref{eq:tcoll} and using an initial planetesimal mass of $281\, \textrm{M}_{\oplus}$, following the results from Simulation \#4 of Paper 1. The dotted line marks where the collision timescale equals the time of the simulation. Even though the collision timescale for one planetesimal can be as low as $10^4\, \textrm{yr}$, it is still roughly an order of magnitude larger than the actual simulated time, meaning that planetesimal-planetesimal collisions can be safely ignored in our simulations as a source for dust production. }
    \label{fig:collTime_nominal_HLTau}
\end{figure}

The timescale for planetesimal-planetesimal collisions is calculated in the same way as for the Solar System simulations and we use the results from Paper 1 to infer the total mass of planetesimals formed at each gap edge. After $1\, \textrm{Myr}$, $281\, \textrm{M}_{\oplus}$ of planetesimals has formed in Simulation \#4 of Paper 1. We make the assumption that the amount of planetesimals forming at each gap edge is the same, which based on the results of Paper 1 is a reasonable approximation. This yields a collision timescale of $\sim\!\! 10^4\, \textrm{yr}$ in the beginning of the simulation, which then increases roughly linearly with time as the planetesimals spread out across the disc (see Figure \ref{fig:collTime_nominal_HLTau}). Since the collision timescale remains an order of magnitude larger than the actual simulated time throughout the simulation, also for HL Tau we conclude that planetesimal-planetesimal collisions are not a major source for dust production

The amount of mass that has been ablated as a function of semimajor axis and time for the nominal model is presented in Figure \ref{fig:abl_HLTau}. From the left plot we learn that all mass loss occurs interior of the innermost planet, with a peak at $\sim\!\! 7\, \textrm{au}$. In total, 11\% of the initial planetesimal mass has been ablated after $1\, \textrm{Myr}$. This is a much smaller number than what we found in the Solar System simulations, and the main reason for this is that the planetesimals form further out in the disc. In order for a planetesimal forming at $90\, \textrm{au}$ to experience efficient ablation, it needs to be scattered inwards by approximately $80\, \textrm{au}$. Since the planetesimals need to travel a further distance before mass loss can begin to occur, the initial mass loss rate is much lower than in the Solar System simulations, where planetesimals start to ablate efficiently already after a few thousand years. This can also be seen in the top panel of Figure \ref{fig:ml_a_allParam_HLTau}, where the mass loss evolution for all planetesimals in the 10 nominal simulations has been plotted. After $10\, \textrm{kyr}$, the planetesimal with the smallest semimajor axis has still only lost about 30\% of its mass. 

\subsection{Parameter study and mass loss profiles}\label{subsect: HL Tau param}
In this parameter study we choose to vary the composition of the planetesimals, the $\alpha$ value and the initial mass accretion rate $\dot{M}_0$. Changing these disc parameters does not result in a significant change to the collision timescale compared to the nominal simulation. As previously explained, lowering $\alpha$ results in an increased surface density, and lowering $\dot{M}_0$ results in a decreased surface density. This change in the surface density also affects the positions of the icelines. Lowering $\dot{M}_0$ by a factor of 10 (Simulation \#11) does not result in any change of the planetesimal compositions compared to the nominal simulation; however, when $\alpha$ is lowered by a factor of 10, then the planetesimals forming at the gap edge of the outermost planet end up interior of the CO iceline. This gives us an opportunity to study CO ablation, since the planetesimals will consist of both H$_2$O, CO$_2$ and CO ice. 

\begin{figure*}
\centering
    {\includegraphics[width=\columnwidth]{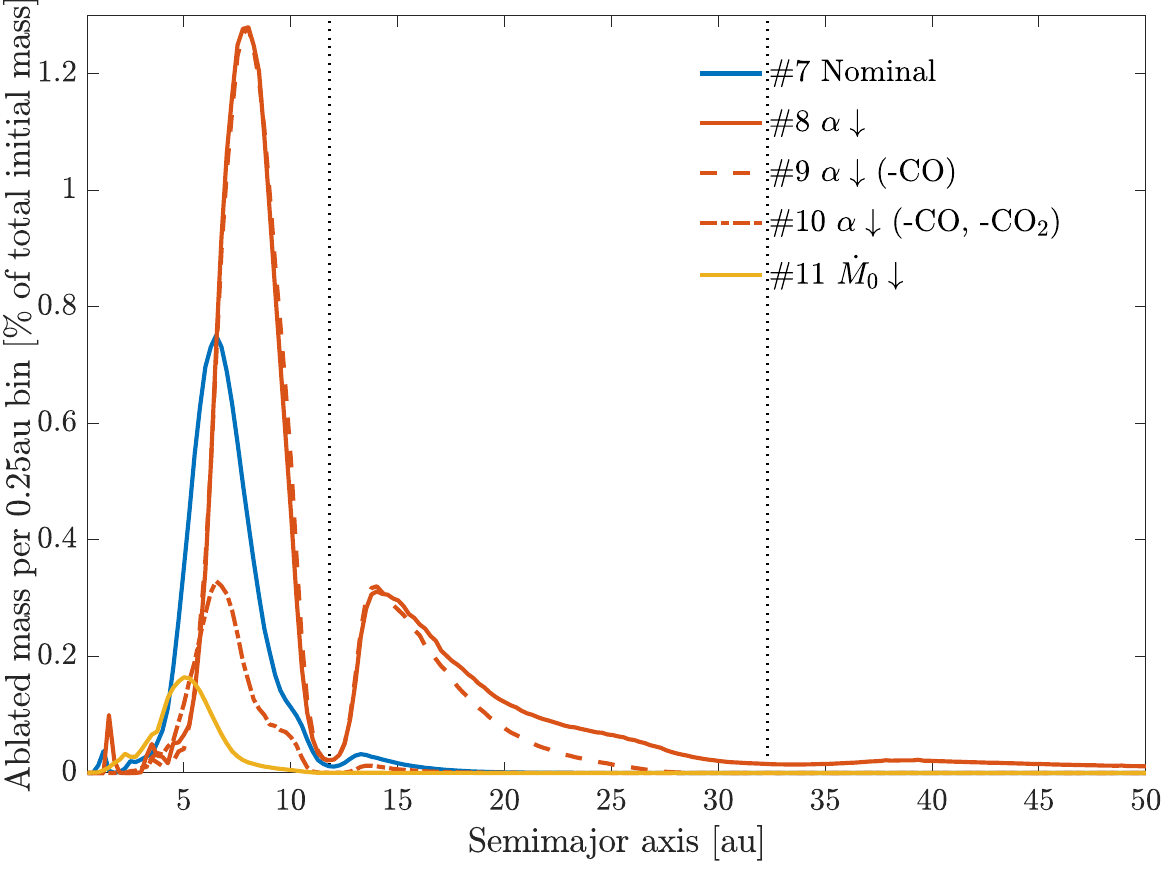}}
    {\includegraphics[width=\columnwidth]{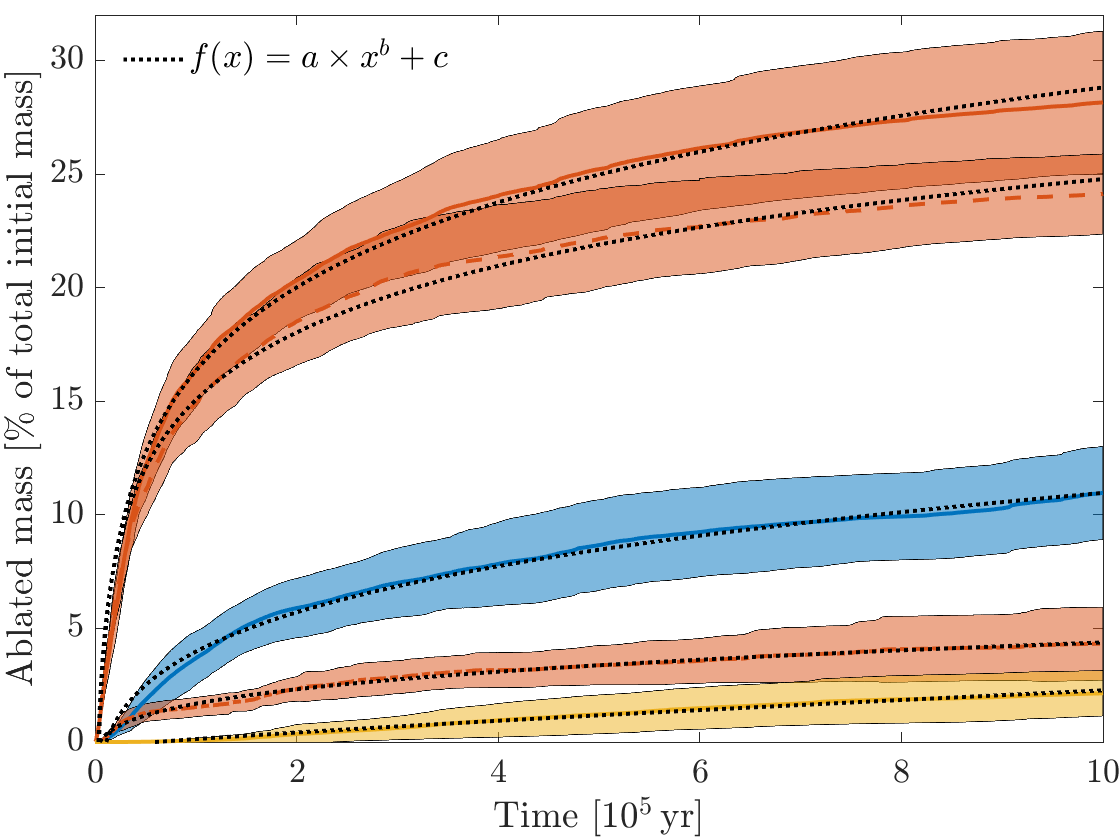}}
\caption{Left plot: distribution of ablated mass across the disc after $1\, \textrm{Myr}$ for all HL Tau simulations. The values on the $y$-axis represent the amount of mass that has been ablated in a $0.25\, \textrm{au}$ semimajor axis bin. The dotted lines mark the semimajor axes of the two innermost planets. In the nominal simulation all mass loss occurs just interior of the innermost planet. When the surface density is increased (\#8) there is also some mass loss occurring further out in the disc. Right plot: the total amount of mass that has been ablated as a function of time, for the same data as in the left plot. The colored lines show the average over the 10 simulations, and the colored region shows the one standard deviation away from this value. The dotted black lines show the best fit to the curve $f(x)$, and the parameters to the fits can be found in Table \ref{table:param HL Tau}. Comparison of Simulation \#8-\#10 shows that most mass loss occurs due to the presence of CO$_2$. }
    \label{fig:abl_HLTau}
\end{figure*}

\begin{figure*}
\centering
\resizebox{\hsize}{!}
{\includegraphics{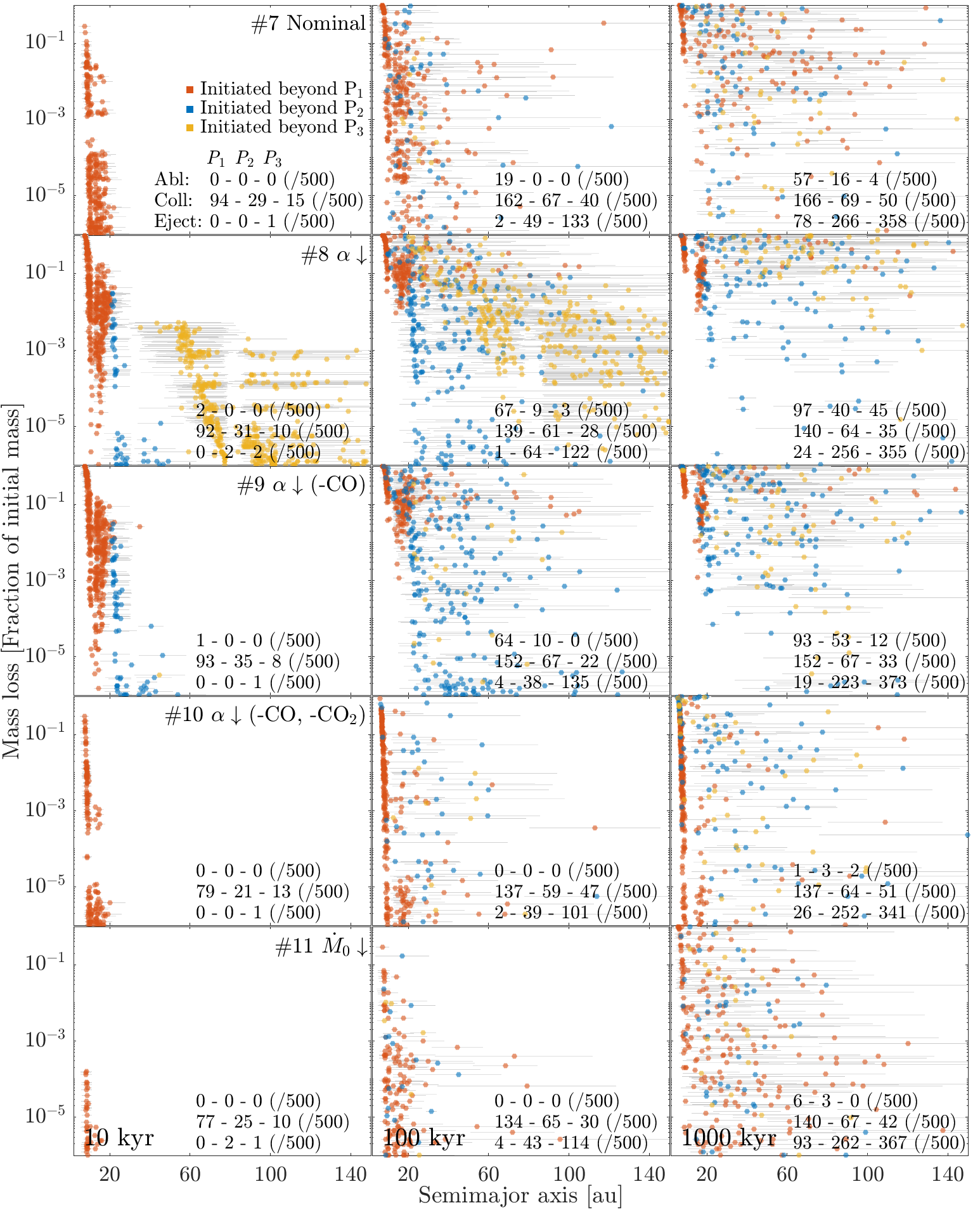}}
\caption{Mass loss versus semimajor axis evolution for all HL Tau simulations. The red dots are for planeteseimals forming at the gap edge of the innermost planet; the blue dots are for planetesimals forming at the gap edge of the middle planet; and the yellow dots are for planetesimals forming at the gap edge of the outermost planet. Ablation is most efficient for planetesimals forming at the gap edge of the innermost planet. Ablation of CO (\#8) causes planetesimals to lose mass far out in the disc. The massive planets deliver strong kicks to the planetesimals which causes many of them to leave the simulation domain.}
    \label{fig:ml_a_allParam_HLTau}
\end{figure*}

In order to see the effect of CO ablation, we perform one simulation with it (Simulation \#8), and one simulation without it (Simulation \#9). Note however that the only planetesimals which will experience CO ablation in Simulation \#8 are those that form at the outermost planetary gap edge. Furthermore, we perform one simulation where we only consider the ablation of water ice (Simulation \#10). The results from these simulations are presented in Figure \ref{fig:abl_HLTau} and \ref{fig:ml_a_allParam_HLTau}. 

When only the ablation of water ice is considered, all mass loss occurs interior to the innermost planet. The total amount of mass that has been ablated after $1\, \textrm{Myr}$ is only about 5\%. When CO$_2$ ablation is added, this value increases to 25\%, and there is significant mass loss also in the region around the gap edge of the innermost planet. Compared to the nominal simulation, the initial mass loss rate is much higher, which is because of the larger surface density of the disc. Finally, when CO ablation is added, the total amount of ablated mass increases by another few percent. Importantly, a small amount of mass loss now occurs as far out in the disc as $100\, \textrm{au}$. This is clearly visible in Figure \ref{fig:ml_a_allParam_HLTau}, where planetesimals forming at the outermost planetary gap edge are losing mass at a slow rate all over the disc.

The effect of decreasing $\dot{M}_0$ is very much the same as in the Solar System simulations: the mass loss rate is lower, and the mass loss occurs closer to the central star. In such a disc, less than 5\% of the planetesimal mass is lost due to ablation. In summary, planetesimals forming at the edges of planetary gaps in the HL Tau system are not as affected by ablation as those forming at the gap edges of Jupiter and Saturn. This is expected, since the planets in the HL Tau system are located much further away from the central star. Nevertheless, if the surface density of the disc is large, the amount of mass loss can still be significant. Most of this mass is released interior of the planets, but some small amount is also some released around the gap edge of the innermost planet. If the surface density of the disc is high enough for CO to be in solid phase where some planetesimals form, these planetesimals can ablate at a slow rate far out in the disc. 

\begin{table}
\centering
\caption{Parameters to the fitted curve $f(x)=a\times x^b+c$ which is shown in the right panel of Figure \ref{fig:abl_HLTau}.}
\label{table:param HL Tau}
\begin{tabular}{llll}
\hline\hline
Run & $a$    & $b$     & $c$     \\ \hline
\textbf{\#7} Nominal   & 10.1  & 0.2282  & -6.14  \\
\textbf{\#8} $\alpha\downarrow$   & 193.1 & 0.0272 & -176.70  \\
\textbf{\#9} $\alpha\downarrow$ (-CO)   & -223.0 & -0.0194 & 238.00  \\
\textbf{\#10} $\alpha\downarrow$ (-CO, -CO$_2$)   & 2.89 & 0.2859 & -1.21  \\
\textbf{\#11} $\dot{M}_0\downarrow$   & 0.4  & 0.8262  & -0.25   \\ \hline
\end{tabular}
\end{table}

\section{Pebble flux due to ablation}\label{sec: the ablated material}
The ablated material consist of a mixture of silicate grains, carbon grains and vaporized ices. For simplicity we assume that all of this material lands on the midplane, which is not a bad approximation given that ablation should be most efficient in the dense midplane of the disc. Depending on where in the disc the ablation occurs, the vaporized ices will either re-condense to form solid ice, or remain in gas-phase. For example, if the ablation occurs in between the H$_2$O and CO$_2$ iceline, then the vaporized H$_2$O ice will re-condense to form solid ice, but CO$_2$ and CO will remain in gas-phase. We assume that re-condensation happens instantaneously. We further assume that all solid material (silicate grains, carbon grains and re-condensed ices) grow to $\sim$mm-sized pebbles with Stokes number $10^{-3}$ directly after the ablation occurs. From Paper 1 we find that $\mu \textrm{m}$-sized grains grow to mm-sizes within $\sim\!\! 10\, \textrm{kyr}$ in the inner region of the disc, so given the timescales under consideration, it is a valid approximation. 

The solid material from ablation gives rise to a flux of pebbles, which can be calculated given the initial planetesimal formation rate at the gap edges, the rate at which these planetesimals are ablating (right panel of Figure \ref{fig:abl_SS} and \ref{fig:abl_HLTau}), and the distribution of the ablated material (left panel of Figure \ref{fig:abl_SS} and \ref{fig:abl_HLTau}). We assume that all planetesimals ablate at the same rate, and that their ablated material has the same distribution, which gives the correct behavior for the population as a whole. The planetesimal formation rate is taken directly from Paper 1 in the case of HL Tau, and calculated using some assumptions in the case of the Solar System (description in section \ref{subsec: abl SS}). This information is combined to give the total mass ablation rate (gas + solids) in each semimajor axis bin and at each timestep of the simulation $\dot{M}_{\textrm{abl}}(r,t)$. We then remove the mass fraction from ices that do not re-condense (using abundances from \citet{Oberg2011} and assuming a chemical composition of Mg$_2$SiO$_4$ for the silicate grains, and C for the carbon grains), and are left with only the solid (pebble) component.

The flux of pebbles in an annulus of with $\Delta r$ is 
\begin{equation}
F(r,t) = \frac{ \dot{M}_{\textrm{abl}}(r,t) }{ 2\pi r \Delta r },
\end{equation}
where $F$ is in units of $\textrm{kg}\, \textrm{s}^{-1}\, \textrm{m}^{-2}$. 
The total mass flux $F_{\textrm{tot}}(r,t)$ as a function of semimajor axis and time is then obtained by numerically integrating $2\pi r F(r,t)$ from $r$ to some $r_{\star}$. Here we make another assumption that the planetary gaps act as hard barriers which completely blocks the flow of pebbles past the gaps. In other words: when calculating the total mass flux interior of the innermost planet, we take the integral from $r$ to the semimajor axis of the innermost planet; similar in between two planets; and when calculating the mass flux exterior to the outermost planet, we take the integral from $r$ to the edge of our ablation array ($20\, \textrm{au}$ for SS, $100\, \textrm{au}$ for HL Tau). Finally, we calculate the corresponding pebble-to-gas surface density ratio using Equation 8-13 from \citet{Johansen2019}. 

\subsection{Solar System}\label{subsec: abl SS}
In order to calculate how much mass is ablated in the Solar System simulations, we first need an estimate for the initial planetesimal formation rate at Jupiter's and Saturn's gap edges. For Jupiter, we simply assume that all pebbles initially located between the semimajor axis of Jupiter and Saturn are turned into planetesimals at Jupiter's gap edge (this is the same assumption as we used for the collision timescale). Using the parameters of the nominal model and a dust-to-gas ratio of 1\%, this results in $16\, M_{\oplus}$. We calculate that it takes $55\, \textrm{kyr}$ for all these pebbles to reach Jupiter's gap edge, using equations for the pebble drift timescale from \citet{Johansen2019}. Assuming a constant drift velocity for all pebbles, this yields an initial planetesimal formation rate of $3\times 10^{-4}\, M_{\oplus}\, \textrm{yr}^{-1}$  at Jupiter's gap edge. This is the rate during the first $55\, \textrm{kyr}$ of the simulation, after which it is zero. In order to keep things simple, we assume the exact same formation rate at Saturn's gap edge. 

\begin{figure}
\centering
    {\includegraphics[width=\hsize]{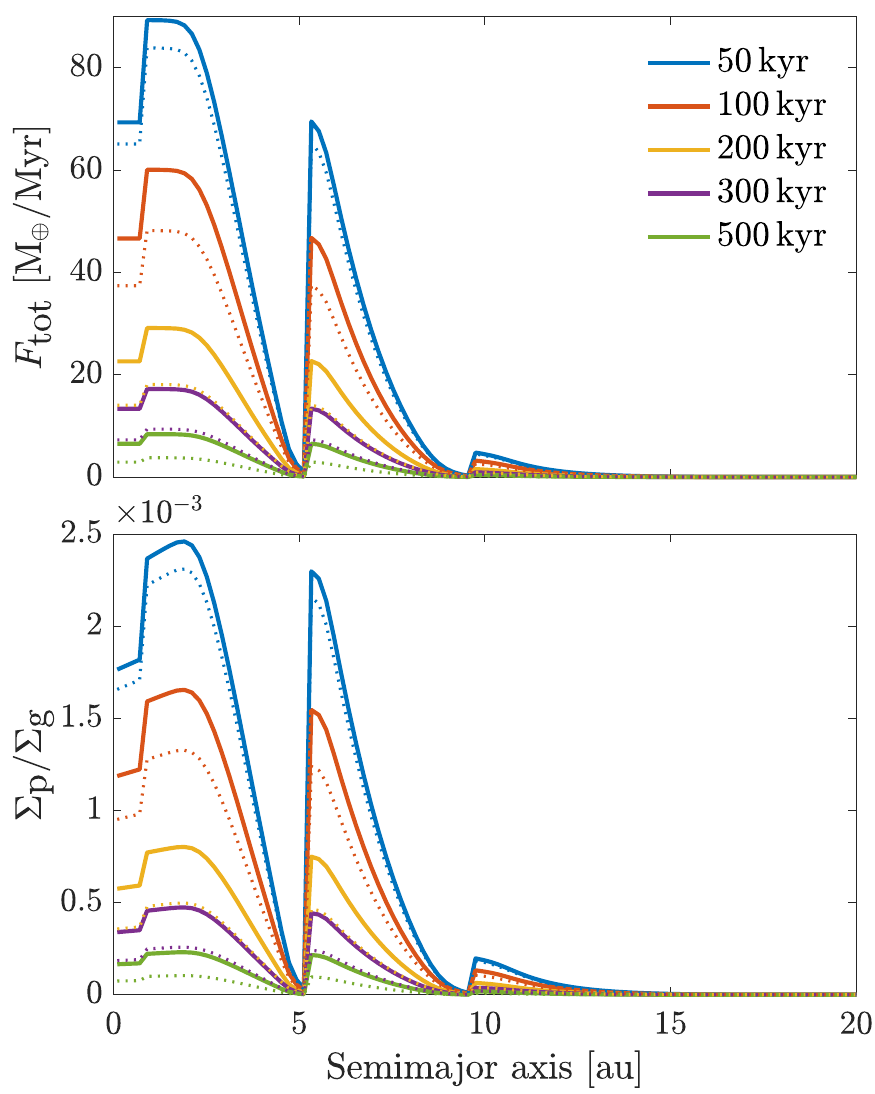}}
\caption{The material which is ablated from the planetesimal surfaces re-condensate and grow to mm-sized pebbles in the disc midplane. In the top plot we show how the mass flux of such pebbles varies as a function of time and semimajor axis for the nominal Solar System simulation. The solid lines take into account the formation and ablation of newer generations of planetesimals forming at the gap edges from this flux of pebbles, and the dotted lines does not. In the bottom panel we show the corresponding pebble-to-gas surface density ratios. }
	\label{fig:flux_sigma_nominal_SS}
\end{figure}

As a first approximation, we neglect any later generations of planetesimals forming at the gap edges from the ablated material. The resulting pebble flux and pebble-to-gas surface density ratio for the nominal model is presented in Figure \ref{fig:flux_sigma_nominal_SS} as dotted lines. The highest pebble flux is obtained at $50\, \textrm{kyr}$, after which it decreases with time. The drop in pebble flux at the H$_2$O iceline is clearly visible. The CO$_2$ iceline is located in the gap just interior of Jupiter, and therefore there is no corresponding drop in the pebble flux.

Since we assumed a hard pebble barrier at the planetary gaps, all pebbles flowing into the gap edge will become trapped. If we use the same assumption as for the initial planetesimal formation rate, all these pebbles will collapse into newer generations of planetesimals. We follow the ablation of these planetesimals, and even later generations, assuming the same ablation rate and distribution of ablated material as for the initial population. This results in an increased pebble flux and surface density ratio compared to our first approximation (see solid lines in Figure \ref{fig:flux_sigma_nominal_SS}). In total about $\sim\!\! 15\, M_{\oplus}$ of pebbles are delivered to $1\, \textrm{au}$ over the course of the simulation. Interior of the water iceline, this amount drops to $\sim\!\! 12\, M_{\oplus}$. The presented fluxes are obtained in a young and massive disc, and would not be present in older and/or low-density discs, thus not preventing the formation of transition discs which are generally old.

\subsection{HL Tau}
\begin{figure}
\centering
    {\includegraphics[width=\hsize]{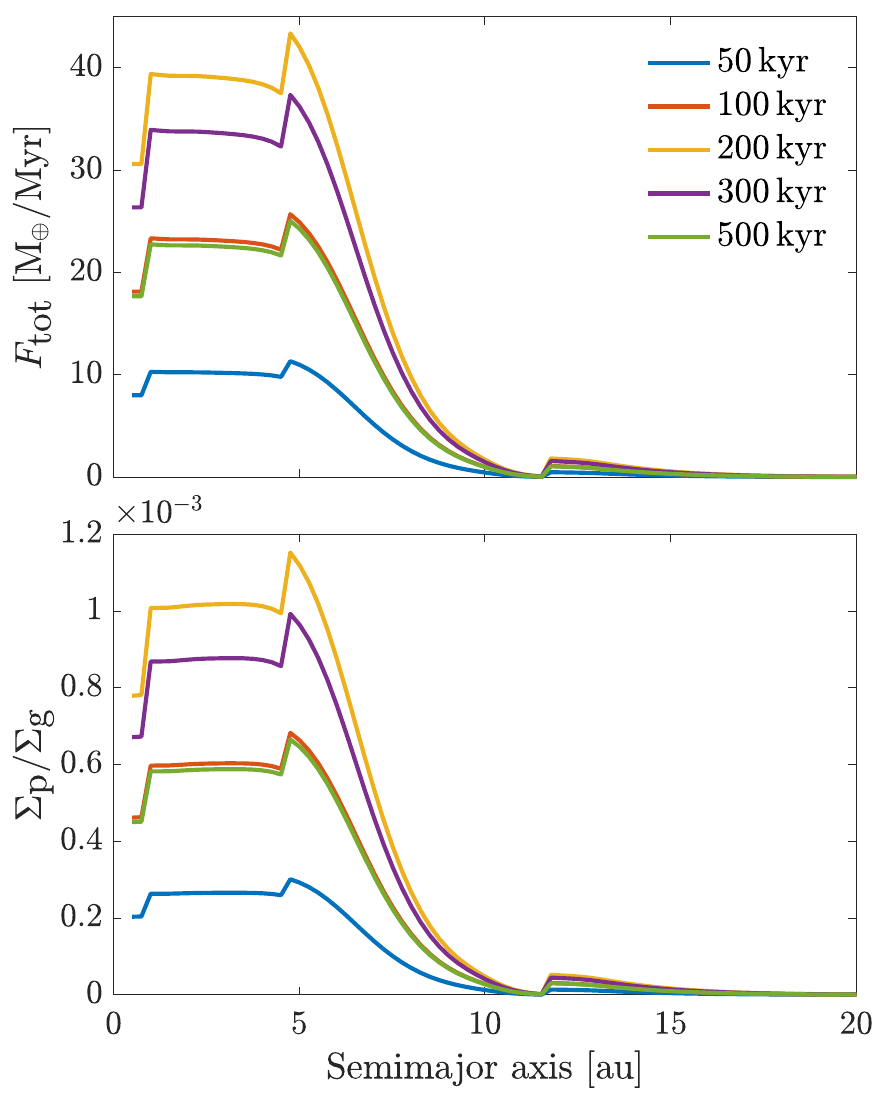}}
\caption{Flux of pebbles (top panel) and pebble-to-gas surface density ratio (bottom panel) as a function of semimajor axis and time for the nominal HL Tau simulation (similar to Figure \ref{fig:flux_sigma_nominal_SS}). Since the large majority of mass ablation occurs interior of the innermost planet, we do not consider the formation of newer generations of planetesimals at the planetary gap edges.}
	\label{fig:flux_sigma_nominal_HLTau}
\end{figure}

In Simulation \#4 of Paper 1 (which has the same parameters as in the nominal simulation) we have $281\, M_{\oplus}$ of planetesimals forming at the gap edges within $1\, \textrm{Myr}$. By using the planetesimal formation rate from this simulation, and combining it with the ablation rate and distribution of ablated mass from this paper, we obtained the amount of mass which is ablated in each semimajor axis bin and at each timestep. The resulting pebble flux and pebble-to-gas surface density ratio is shown in Figure \ref{fig:flux_sigma_nominal_HLTau}. Since the amount of ablated material being deposited beyond the innermost planet is very small, we do not consider the formation of newer generation of planetesimals. 

Due to the relatively low ablation rate in the nominal simulation (only 11\% of the initial planetesimal mass is ablated during $1\, \textrm{Myr}$), the pebble-to-gas surface density ratio only reaches about 0.1\%. The maximum pebble flux is about half that of the Solar System, $40\, M_{\oplus}\, \textrm{Myr}^{-1}$, and the total integrated mass flowing past $1\, \textrm{au}$ is $25\, M_{\oplus}$. In the Solar System simulations, most ablation occurs within the first $50\, \textrm{kyr}$, after which the ablation rate drops quickly. In the HL Tau simulations, ablation happens on a longer timescale, and the rate does not drop as quickly after the peak at $200\, \textrm{kyr}$. Because of this the flux of pebbles remains high for a much longer time; however, we emphasize that these results are for a non-evolving disc and in an evolving disc the pebble flux would decrease with time along with the gas surface density.

\subsection{The asteroid dichotomy}
In this work we have used a simple model of the Solar System with only two planets, Jupiter and Saturn, where neither planet nor disc properties evolve with time. Due to these simplifications, comparisons to Solar System data and observations remain conceptual. We nevertheless comment on how pebble drift due to ablation could fit into our understanding of the accretion history of the inner Solar System, specifically the isotopic dichotomy among the parent bodies of iron meteorites (irons) and chondritic meteorites (chondrites). 

Meteorites collected on Earth can be broadly classified as either carbonaceous (CC) or noncarbonaceous (NC) based on their distinct isotope compositions (\citealt{Warren2011,Kruijer2017}). This isotopic dichotomy, which could reflect either an infall of material with two compositions \citep{Jacquet2019} or a temperature dependent destruction of presolar grains (\citealt{Trinquier2009,Schiller2018}), hints towards the formation of meteorite parent bodies from two distinct and spatially separated reservoirs \citep{Kruijer2017}. Since NC chondrites derive from dry bodies, whereas CC chondrites show hydration features, it is generally envisioned that the NC population represents inner Solar System bodies while members of the CC group would have formed further out.

Iron meteorites are fragments of cores from melted and differentiated planetesimals which accreted within $1\, \textrm{Myr}$ after the formation of CAIs (calcium–aluminum-rich inclusions) \citep{Kruijer2017}. Chondrites are fragments of non-molten and non-differentiated planetesimals which accreted around $2-4\, \textrm{Myr}$ after CAI formation \citep{Kita2012}. The parent bodies of NC and CC irons accreted $\sim 0.4\, \textrm{Myr}$ and $\sim 0.9\, \textrm{Myr}$ after CAI formation, respectively. Since the two populations have distinct compositions they must have remained separated also their after formation. Similarly, the NC and CC reservoirs must have been separated at the time of formation of the NC chondrite parent bodies and CC chondrite parent bodies at $\sim 2\, \textrm{Myr}$ and $3-4\, \textrm{Myr}$ after CAIs, respectively. In the context of pebbles drift, such constraints imply that pebbles bearing a CC isotopic signature either never entered the inner Solar System region where the parent bodies of NC irons and chondrites formed, or, if they did, were not efficiently accreted by the NC population of objects. Below we discuss these issues in the context of our results.

The formation of Jupiter likely begun early and resulted in gap-opening in less than $1\, \textrm{Myr}$ \citep{Kruijer2017}. At this point planetesimals should have started to form at its gap edge. As shown in our simulations, a large fraction of these planetesimals would be scattered into the inner Solar System. While the disc is young and massive a large fraction of these planetesimals become ablated, resulting in a flux of pebbles with (possibly) CC composition both interior and exterior of Jupiter's gap. Given the expected decrease in disc density with time, this process should have stopped to be relevant at the time of formation of the NC chondrite parent bodies at $\sim 2\, \textrm{Myr}$ after CAIs. However, since the NC iron parent bodies formed as early as $\sim 0.4\, \textrm{Myr}$ after CAIs, they should have been in place at the time Jupiter reached the pebble isolation mass and started to scatter planetesimals inside of its orbit. Something must thus have prevented the NC iron parent bodies from accreting the pebbles produced by the early ablation of planetesimals originating from beyond Jupiter’s orbit. This process was likely the early formation of planetary embryos interior of Jupiter's orbit, which caused the inclinations and eccentricities of the planetesimals to be excited, thereby disconnecting them from the pebbles drifting through the midplane \citep{Schiller2018}. As long as the planetesimals remain excited, they would be unable to efficiently accrete pebbles (e.g. \citealt{Levison2015,Liu2019,Johansen2015}) and should thus have preserved their birth compositions. In general, the described pebble flux was present at a time in the Solar System so early that there is likely no record of it.

\section{Planet-planetesimal collisions}\label{sect: collisions}
In the Solar System simulations, about 1-2\% of all planetesimals suffer a collision with a planet. The majority of these collisions are between Jupiter and planetesimals formed at Jupiter's gap edge. Only about 20\% of the collisions are between a planet and a planetesimal not formed at that planet's gap edge. Following the assumption made in Section \ref{subsec: abl SS}, and assuming that the collisions occur before any mass loss has taken place, the total planetesimal mass colliding with Jupiter is about half an Earth mass. Depending on how high up in the atmosphere this material is deposited, this could be relevant for the composition of Jupiter's atmosphere (e.g. \citealt{ShibataIkoma2019}).

Collisions between planets and planetesimals occur much more frequently in the HL Tau system, where about 15-20\% of all planetesimals suffer such a collision. Due to the larger separation between the planetary orbits, close to all collisions are between a planet and a planetesimal formed at that planet's gap edge. If a large fraction of the colliding planetesimal mass is accreted high up in the planet's atmospheres, this would certainly be relevant for the atmosphere's compositions. 

The collision algorithm employed in this work is rather simple, and by making it more advanced the frequency of collisions could certainly change. We use a direct collision search where the radius of the planet is determined by assuming a constant density, but in reality planets that are accreting gas have an inflated envelope with a significantly larger radius. Thus the capture radius used in this paper is significantly smaller than it would be in real life, resulting in a smaller number of collisions. The initial formation location of the planetesimals relative to the planet (discussed in Appendix \ref{Appendix: gapwidth}) could also have an effect on the collision frequency. Another process which has been ignored in this paper but would likely lead to more collisions is planetary migration \citep{Pirani2019,CarterStewart2020}.

\section{Conclusions and future studies}\label{setc: conclusion}
In this work we study the evolution of planetesimals formed at planetary gap edges. To this end we perform a suite of N-body simulations, and consider two planetary systems: the Solar System with Jupiter and Saturn, and a system inspired by HL Tau with 3 planets. We then follow the evolution of planetesimals initiated at the gap edges, where the planetesimals are further subjected to gas drag and mass loss via ablation. We assume that the mass which is ablated from the planetesimals will re-condense to form pebbles in the disc midplane and calculate the corresponding pebble flux.

We reached the following conclusions regarding the main questions posed in the introduction section:  
\begin{itemize}
\item[1.] To what extent do gravitational interactions with the forming planets redistribute the planetesimals formed at their gap edge? The close proximity to the gap-opening planets results in large orbital excitations, causing the planetesimals to leave their birth location soon after formation. Within 10 orbital periods the semimajor axes of the planetesimals have diffused over several au, and after a few hundred orbital periods the planetesimals are spread out across the entire disc. If the gap-opening planets are massive, as in the HL Tau system, $\sim\!\! 40\%$ of the planetesimals become ejected from the system within a million years. These planetesimals could potentially be circularized by galactic tides and end up in the Oort cloud \citep{BrasserMorbidelli2013}. The ejection efficiency generally increases with time as the disc density decreases and the planetary masses increases. If the planetary gaps were to be much wider than assumed in this paper, resulting in formation locations further from the planets, the scattering phase would be slower and delayed.
\item[2.] Can the frictional heating of planetesimals on eccentric orbits drive the production of a significant amount of dust through surface ablation? Planetesimals on eccentric orbits interior of $\sim\!\! 10\, \textrm{au}$ experience efficient ablation, because of this there are no planetesimals entering the innermost disc region. In the nominal Solar System simulation with $\dot{M}_0=10^{-7}\, M_{\odot}\, \textrm{yr}^{-1}$, $\alpha=10^{-2}$, and Jupiter and Saturn at 30\% of their current masses, 70\% of the initial planetesimal mass has been ablated after $500\, \textrm{kyr}$. Planetesimals formed at Jupiter's gap edge lose about two times more mass than planetesimals formed at Saturn's gap edge. The ablation rate is significantly lower in the HL Tau system due to the larger planetary orbits, and only 11\% of the initial planetesimal mass has been ablated after $1\, \textrm{Myr}$ for the same disc parameters. Planetesimals that contain CO$_2$ ice ablate much more efficiently than planetesimals that contain only H$_2$O ice, and planetesimals that further contain CO ice lose a small amount of mass in the furthest regions of the disc. Given the high dependency on the disc density, ablation is only expected to be important in relatively young and massive discs.
\item[3.] How common are collisions between pairs of planetesimals and between planetesimals and planets? We estimated the timescale for planetesimal-planetesimal collisions in our simulations, and we found that it is at least an order of magnitude longer than the actual simulated time. In other words, each planetesimal suffers a low risk of colliding with another planetesimal. About 1-2\% of all planetesimals in the Solar System simulations will collide with either Jupiter or Saturn, while for the HL Tau system this value is about ten times higher. Depending on where in the atmosphere the material from these planetesimals are deposited, this could be relevant for the atmospheric compositions.
\end{itemize}

The material which is ablated from the planetesimal surfaces consist of a mixture of silicate grains, carbon grains and vaporized ices. A large fraction of these vaporized ices re-condense to form solid ice. The solid material quickly grows to millimeter-sized pebbles in the disc midplane, drift towards the star and gives rise to a flux of pebbles. In the Solar System there is a flux of pebbles produced by planetesimal ablation both interior to and exterior to Jupiter's gap edge. Exterior of Jupiter, this pebble flux could give rise to newer generations of planetesimals. The total integrated mass that reaches $1\, \textrm{au}$ is $\sim\!\! 15\, M_{\oplus}$ in the nominal Solar System simulation, and $\sim\!\! 25\, M_{\oplus}$ in the nominal HL Tau simulation. The pebble flux is expected to drop with time as the density in the disc decreases and ablation becomes less important.

In this work, as well as our previous paper \citet{Eriksson2020}, we have focused on studying the formation and fate of planetesimals formed at stationary planetary gap edges, but in reality planets migrate. For the formation of planetesimals, this should result in that planetesimals form in a larger region of the disc \citep{ShibaikeAlibert2020}. The size of this region depends on how far the planet migrates between the onset of planetesimal formation, and the halt of migration when the gap becomes very deep. If planetesimals are forming while the migration timescale is smaller than or comparable to the dynamical timescale, it could result in larger excitations of the planetesimal orbits, and more planetesimal-planetesimal collisions \citep{CarterStewart2020}. The effects of planetary migration will be investigated in a follow-up study.

\begin{acknowledgements}
L.E. and A.J. are supported by the Swedish Research Council (Project Grant 2018-04867). T.R. and A.J. are supported by the Knut and Alice Wallenberg Foundation (Wallenberg Academy Fellow Grant 2017.0287). A.J. further thanks the European Research Council (ERC Consolidator Grant 724 687-PLANETESYS), the G\"{o}ran Gustafsson Foundation for Research in Natural Sciences and Medicine, and the Wallenberg Foundation (Wallenberg Scholar KAW 2019.0442) for research support. The computations were performed on resources provided by the Swedish Infrastructure for Computing (SNIC) at the LUNARC-Centre in Lund, and are partially funded by the Royal Physiographic Society of Lund through grants. Simulations in this paper made use of the REBOUND code which is freely available at http://github.com/hannorein/rebound.
\end{acknowledgements}

\bibliographystyle{aa} 
\bibliography{ref} 

\appendix

\section{No ablation simulation}\label{Appendix: noAbl}

\begin{figure*}
\centering
    {\includegraphics[width=.7\hsize]{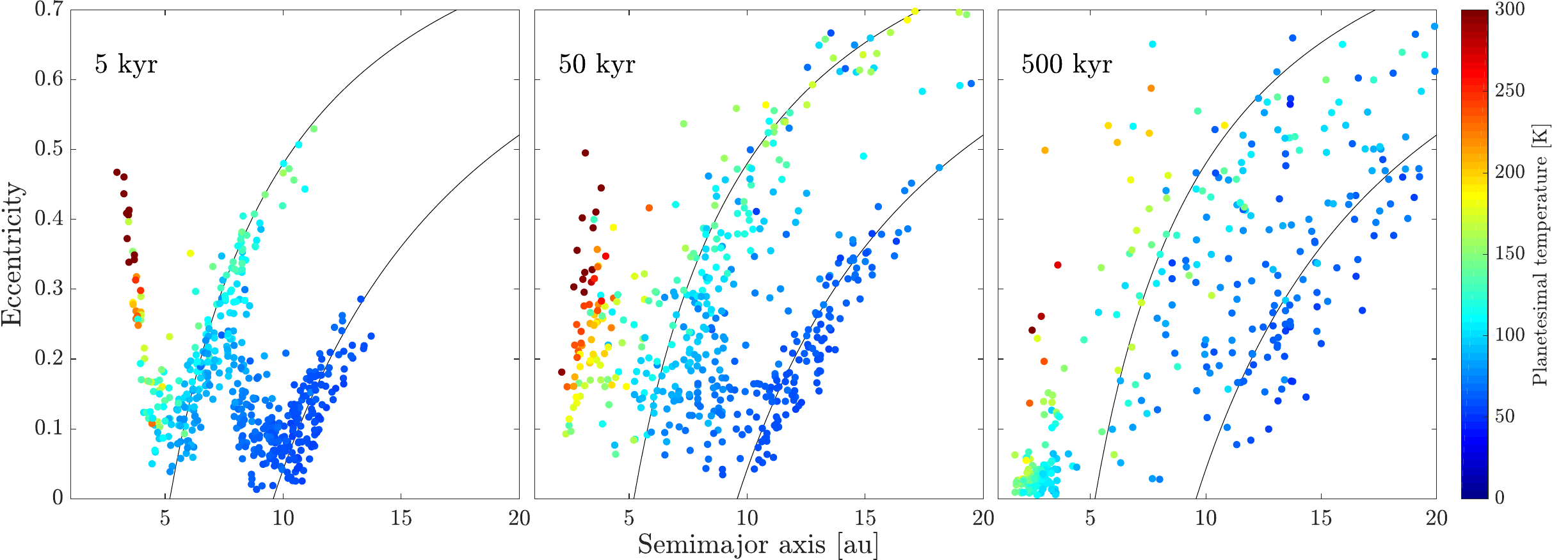}}
    {\includegraphics[width=.7\hsize]{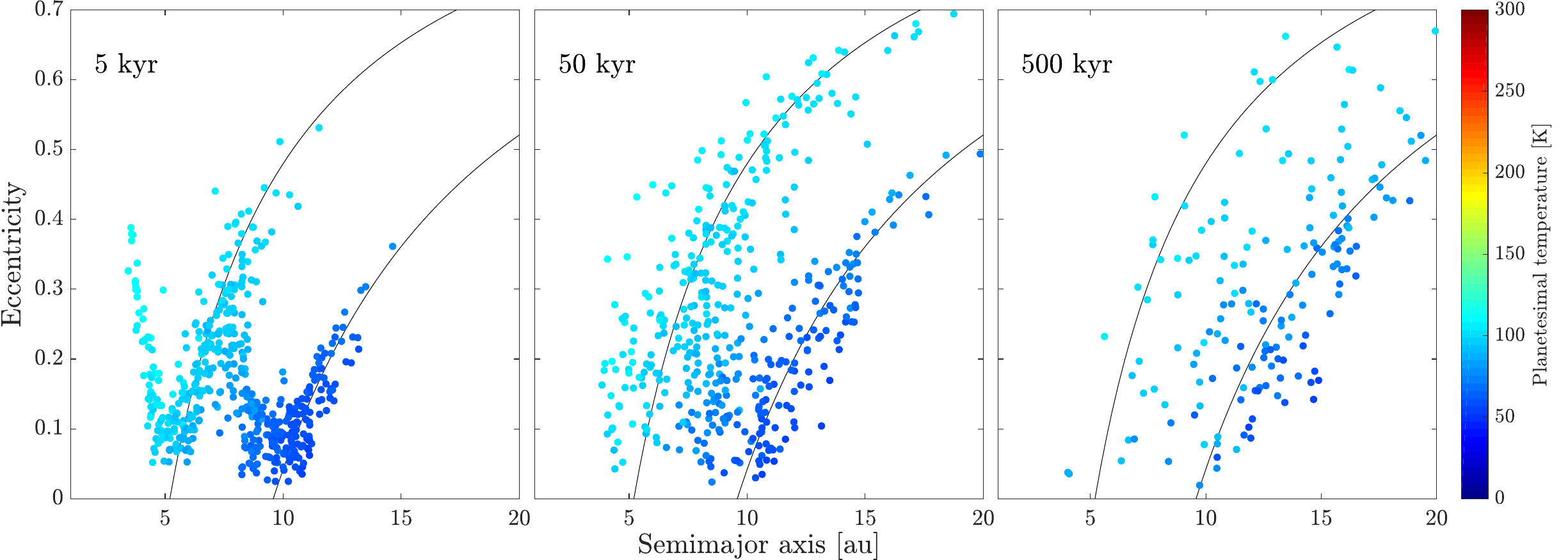}}
\caption{The plots show a comparison of nominal Solar System simulations without surface ablation (top panels), and with surface ablation (bottom panels). The temperature range in the plots is set to $0-300\, \textrm{K}$ to allow for easy comparison, the hottest planetesimals in the top panels have surface temperatures of $443\, \textrm{K}$, $362\, \textrm{K}$ and $281\, \textrm{K}$ respectively going from left to right. The high temperatures obtained in the non-ablation simulations are because of the lack of cooling due to the release of latent heat of vaporization. The population of planetesimals with low eccentricities and short periods in the non-ablation simulations have been circularized by gas drag, this population do not exist in the ablation simulations since they are removed by ablation before they become circularized.}
    \label{fig:e_a_T_noAbl}
\end{figure*}

In Figure \ref{fig:e_a_T_noAbl} we show a comparison of simulations with and without surface ablation. When surface ablation is neglected, there is no cooling due to latent heat of vaporization (last term in equation \ref{eq:Tpl}). In regions where surface ablation would anyway have been inefficient, that is in regions where the surface temperatures are low, this cooling term is very small and thus there is no significant difference between the surface temperatures in the two simulations. This is the case for planetesimals located in Saturn's scattered disc and beyond. For planetesimals located in Jupiter's scattered disc and closer to the star, the surfaces can reach several hundred degrees higher temperatures than they do in simulations including ablation. If an ablating planetesimal in this region were to suddenly stop ablating for some reason, it's surface temperature would rapidly increase.

In runs without ablation the effect of eccentricity damping by gas drag in the innermost disc becomes visible. With time the eccentricities of planetesimals close to the star begin to decrease, and after $500\, \textrm{kyr}$ a population of planetesimals with small and close-to circular orbits emerges. This population does not appear in the ablation simulations, since planetesimals become completely ablated before they are circularized. 

The disc parameters of the nominal simulation correspond to a young and relatively massive disc. The surface density of discs decreases with time, meaning that both ablation and circularization by gas drag will become less efficient. The difference between the two simulations will thus be less prominent in older discs, and just like ablation it will occur closer to the star. This means that as the disc evolves with time and ablation becomes inefficient, planetesimals should eventually survive in the asteroid belt region. Finally we would like to highlight that ablation is a process which occurs in the presence of a major gas component, and it could not occur for planetesimals with similar surface temperatures in a system without gas (e.g. comets).

\section{Comments on crust formation}\label{Appendix: crust formation}
In this work we have assumed that solid grains embedded in the ice are released along with the vaporized ices. In a scenario where the outflow of vapor is too small to carry along the solid grains, one might instead expect that a dry dust mantle builds up. Once this mantle has become thick, it would severely limit further mass loss. Here we present a few arguments for why this should not be the case for the efficiently ablating planetesimals in our simulations. 

In Appendix \ref{Appendix: noAbl} we show that if an efficiently ablating planetesimal suddenly stops ablating, then the lack of cooling due to latent heat of vaporization causes the surface temperatures to increase by up to several hundred degrees. In other words, if a dust mantle starts to build up and begins to limit the amount of mass loss via ablation, the surface temperature quickly increases. A similar effect has been studied in the case of comets, although the heating source is different (\citealt{Orosei1995,Coradini1997,Prialnik1997}). The main differences are that: (1) planetesimals can reach much higher surface temperatures than comets if the gas densities are high, which goes against the sustainability of a dust mantle; and (2) the strong headwind around the planetesimals should blow of the dust that is lifted up (whereas dust can fall back onto comets). The formation of dust mantles in the case of planetesimals is thus more difficult than in the case of comets, and even for comets it is not a process which always occurs \citep{Prialnik1997}.

The process described above relies on the presence of a significant gas component, which is the case for planetesimals on short orbits in young discs. Further away from the star where the gas density is lower and ablation less efficient, this process is much less dramatic and a dust mantle might be allowed to form; however, since planetesimals in this region do not ablate very efficiently to begin with, the effect should not modify our results significantly. The production of a dust mantle will also get easier with time, as the gas density decreases and ablation becomes less and less efficient.

In order to further validate our method we have performed a calculation to check whether the escaping gas from planetesimals is strong enough to carry mm-sized pebbles along with it. This is assumed to be the case when the drag force exerted on a pebble by the outflowing gas is stronger than the gravitational pull of the planetesimal (we ignore the centrifugal correction on the gravity of the planetesimal). Because of the strong headwind in the disc, any pebble that is lifted up from the surface will be immediately carried away by the headwind. As previously mentioned, this is in contrast with the activity of airless bodies such as comets for which dust grains should fall back if their velocity is smaller than the escape velocity of the body.

The initial vertical velocity of a pebble at the surface is calculated as
\begin{equation}
v_0 = v_{\textrm{th}} - \frac{G M_{\textrm{pl}} t_s}{R_{\textrm{pl}}^2}.
\end{equation}
Here $v_{\textrm{th}}$ is the outflow speed of the gas molecules, and we perform the calculation for CO$_2$ molecules. The density of the outflowing gas is calculated as
\begin{equation}
\rho = \frac{ \dot{m}_{\textrm{abl}} }{ A_{\textrm{pl}} v_{\textrm{th}} },
\end{equation}
where $A_{\textrm{pl}}$ is the surface area of the planetesimal.
The relative velocity between the pebble and the gas, which is required to calculate the stopping time, is then simply
\begin{equation}
v_{\textrm{rel}} = v_0 - v_{\textrm{th}} = \frac{G M_{\textrm{pl}} t_s}{R_{\textrm{pl}}^2}.
\end{equation}
Since $t_s$ depends also on $v_{\textrm{rel}}$ this equation has to be solved using a bisection method. 

\begin{figure*}
\centering
    {\includegraphics[width=.7\hsize]{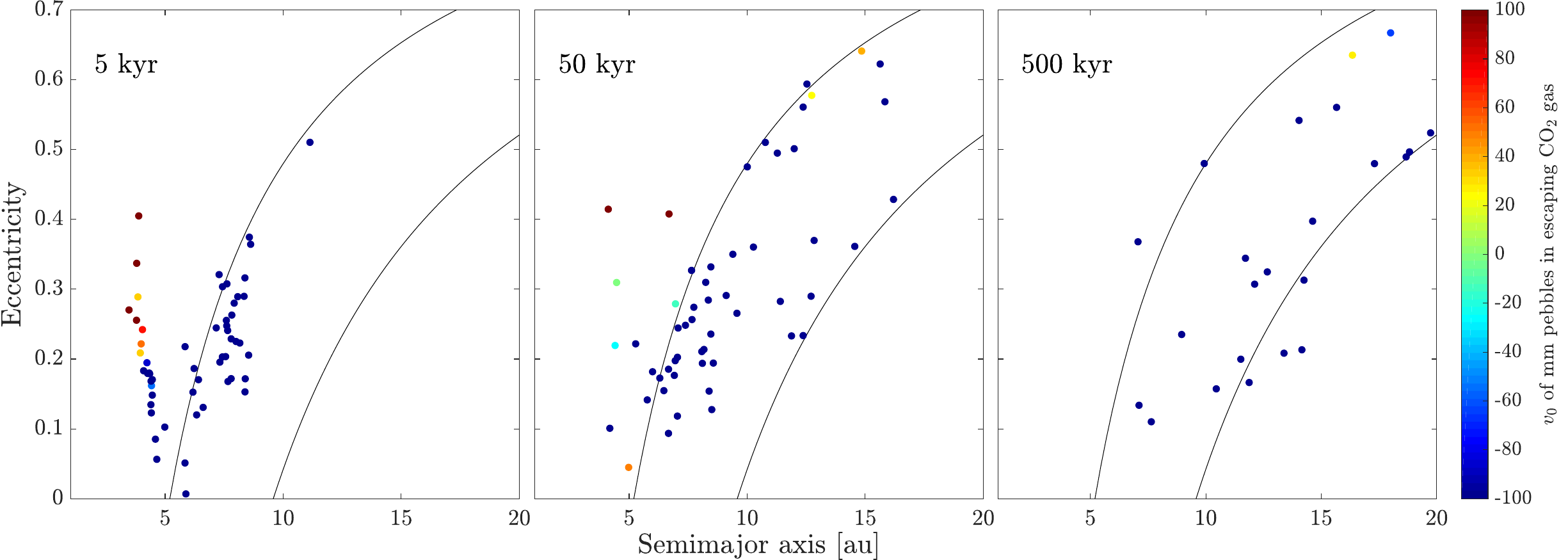}}
\caption{Initial vertical velocity for mm-sized pebbles in the outflowing CO$_2$ gas caused by surface ablation. The calculation is done for planetesimals around perihelion in the nominal Solar System simulation. Pebbles are released along with the gas in the case of high eccentric planetesimals interior of Jupiter's orbit. Lighter gas molecules, smaller pebble sizes and increased surface temperatures due to a lack of cooling from release of latent heat of vaporization would result in more pebbles being released in this plot.}  
    \label{fig:e_a_v0}
\end{figure*}

The results from this calculation is presented in Figure \ref{fig:e_a_v0} for planetesimals close to perihelion in the nominal Solar System simulation. In this plot a pebble is being ejected from the planetesimal when it's initial vertical velocity is larger than 0. The results show that pebbles are lifted up by the escaping gas in the case of the most efficiently ablating planetesimals. As explained earlier, ablating planetesimals that start to build up a dust mantle would obtain significantly increased surface temperatures. The corresponding initial vertical velocity of pebbles on those planetesimals would then increase, leading to much more released pebbles than what's suggested by the plot. The above calculation is performed for CO$_2$ molecules. Using H$_2$O or CO molecules instead results in higher pebble velocities, provided the ablation rate is the same. The pebble size also affects the calculation, with larger pebbles being more difficult to lift off the surface (e.g. \citealt{Orosei1995}).

\begin{figure}
\centering
    {\includegraphics[width=\hsize]{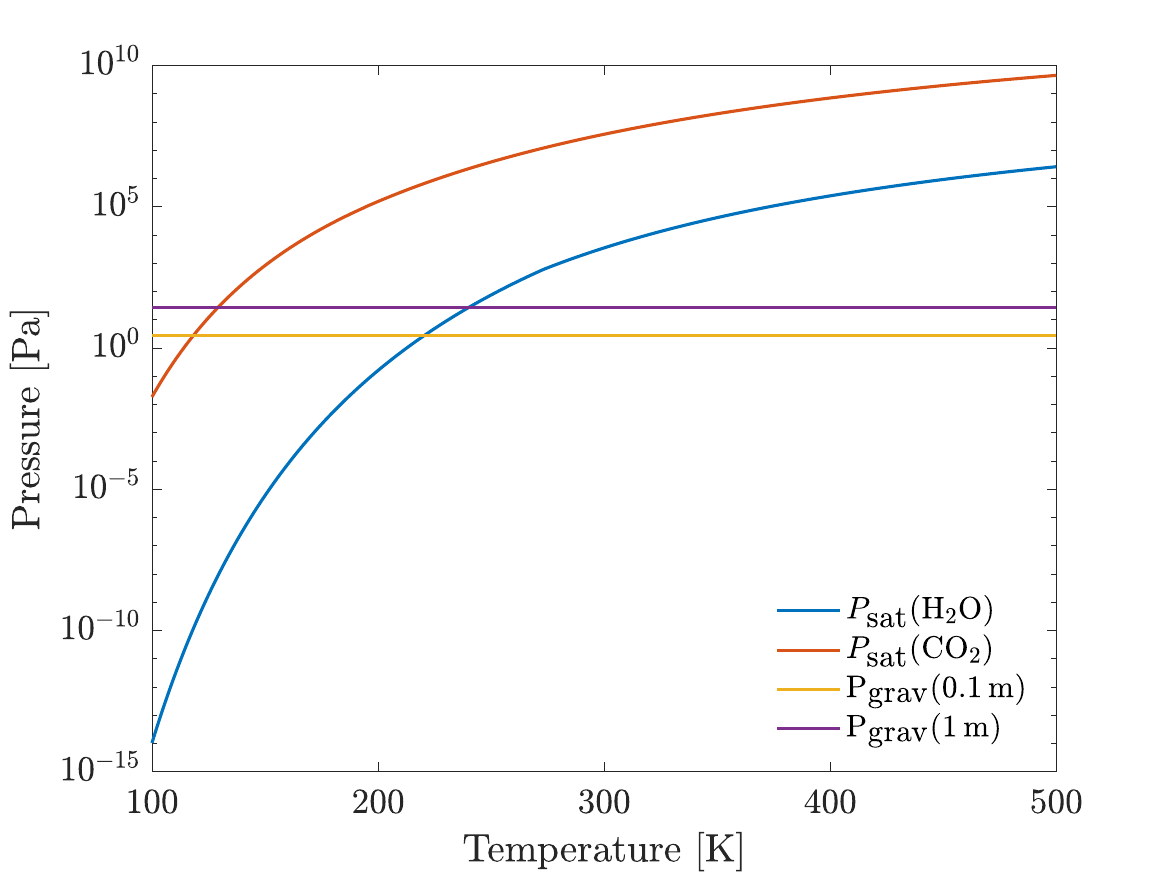}}
\caption{ Comparison between the saturated vapor pressure and the gravitational pressure at $0.1\, \textrm{m}$ and $1\, \textrm{m}$ below the surface of a $100\, \textrm{km}$-sized planetesimal at different temperatures and for different molecules. The temperature down to $1\, \textrm{m}$ below the surface is assumed to be the same as on the surface. The temperature needs to be about twice as high when considering H$_2$O molecules than when considering CO$_2$ molecules in order for the vapor pressure to win over the gravitational pressure. } 
    \label{fig:P_T_1mcrust}
\end{figure}

Finally, we ask ourselves whether a potential crust would have survived the pressure from the vapor sublimated underneath its surface. To answer this we perform a simple calculation where we compare the saturated vapor pressure with the gravitational pressure at $0.1\, \textrm{m}$ and $1\, \textrm{m}$ below the surface of a $100\, \textrm{km}$-sized planetesimal, assuming constant density, hydrostatic equilibrium and that the temperature down to $1\, \textrm{m}$ below the surface is the same as on the surface. The result of this calculation is presented in Figure \ref{fig:P_T_1mcrust}. When considering CO$_2$ vapor, a temperature of about $120-130\, \textrm{K}$ is required to blow off the crust, for H$_2$O a temperature of $220-240\, \textrm{K}$ is required. Comparison with the top panel of Figure \ref{fig:e_a_T_noAbl} shows that these temperatures are obtained for planetesimals sitting inside and interior of Jupiter's scattered disc in the case of CO$_2$ vapor, and for highly eccentric planetesimals interior of Jupiter's orbit in the case of H$_2$O vapor. Any crust that forms in these regions of the disc is thus likely to be blown off. If a crust were to form further out in the disc on the other hand, it would likely remain and limit any further ablation from the planetesimal surface. \textbf{However, note that the formed crust is likely to be porous with a density smaller than what is assumed for this calculation, so the temperatures required to blow off the crust would be lower than suggested by Figure \ref{fig:P_T_1mcrust}.}  

\section{Saturated vapor pressure}\label{Appendix: Psat}

\begin{figure}
\centering
    {\includegraphics[width=\hsize]{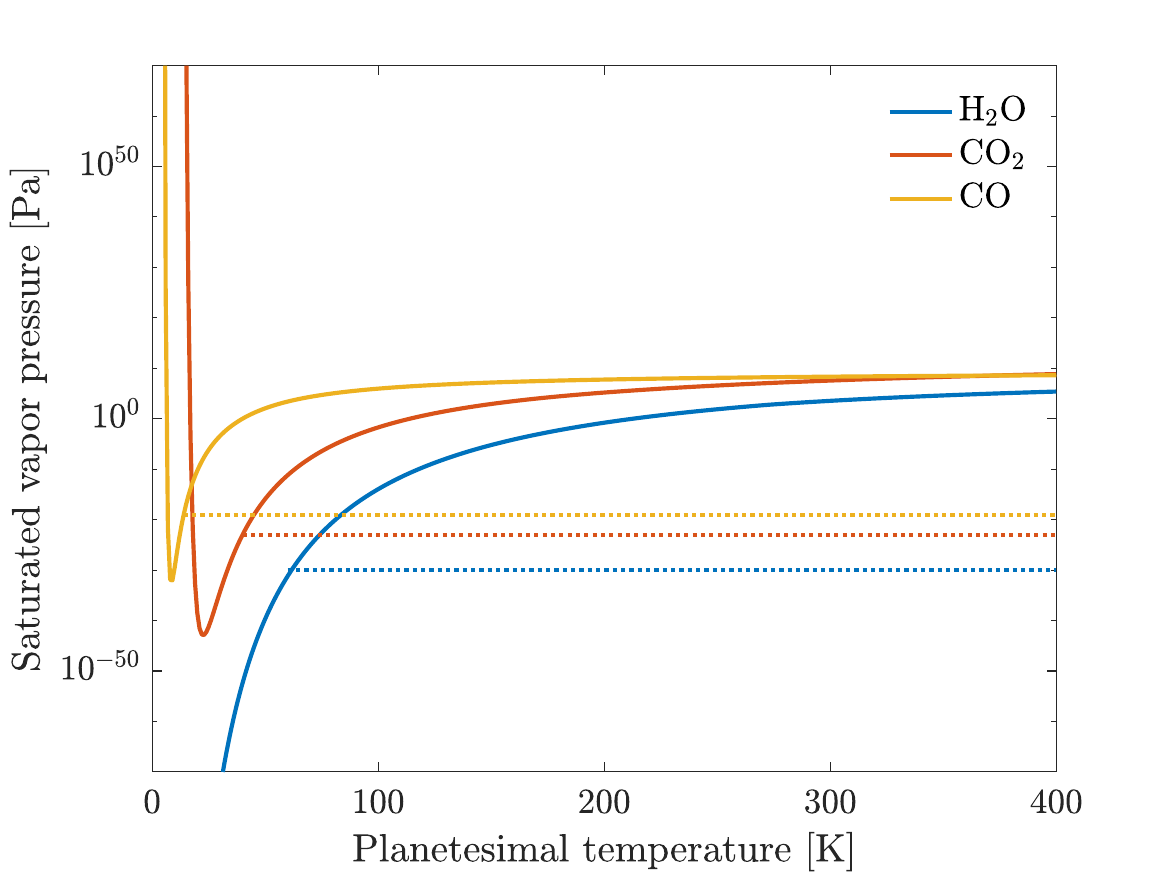}}
\caption{Saturated vapor pressure as a function of planetesimal surface temperature for H$_2$O, CO$_2$ and CO ice. The dotted lines mark our floor values, which are used for numerical reasons.}
    \label{fig:Psat_Tpl}
\end{figure}

Figure \ref{fig:Psat_Tpl} shows the saturated vapor pressure as a function of the planetesimal surface temperature.
The saturated vapor pressure is calculated using polynomial expressions from \citet{FraySchmitt2009}. These expressions are valid in the following temperature range: $0-273.16\, \textrm{K}$, $40-216.58\, \textrm{K}$ and $14-68.1\, \textrm{K}$ for H$_2$O, CO$_2$ and CO ice, respectively. Looking at Figure \ref{fig:Psat_Tpl}, there is a smooth extrapolation to temperatures above the stated temperature range for all ices. This is not the case below the temperature range. For CO$_2$ and CO ice the polynomial turns upwards at low temperatures; therefore, we introduce floor-values at $40\, \textrm{K}$ and $14\, \textrm{K}$, respectively. For H$_2$O the polynomial heads to minus infinity as the temperature decreases; therefore, we introduce a floor-value at $60\, \textrm{K}$ to prevent numerical issues. Additional tests show that lowering this floor-value does not affect the results. Since there are no issues at the upper temperature range, we do not introduce any similar constraints there. 
 
\section{Mass ablation rate}\label{Appendix: ablRate}

\begin{figure}
\centering
    {\includegraphics[width=\hsize]{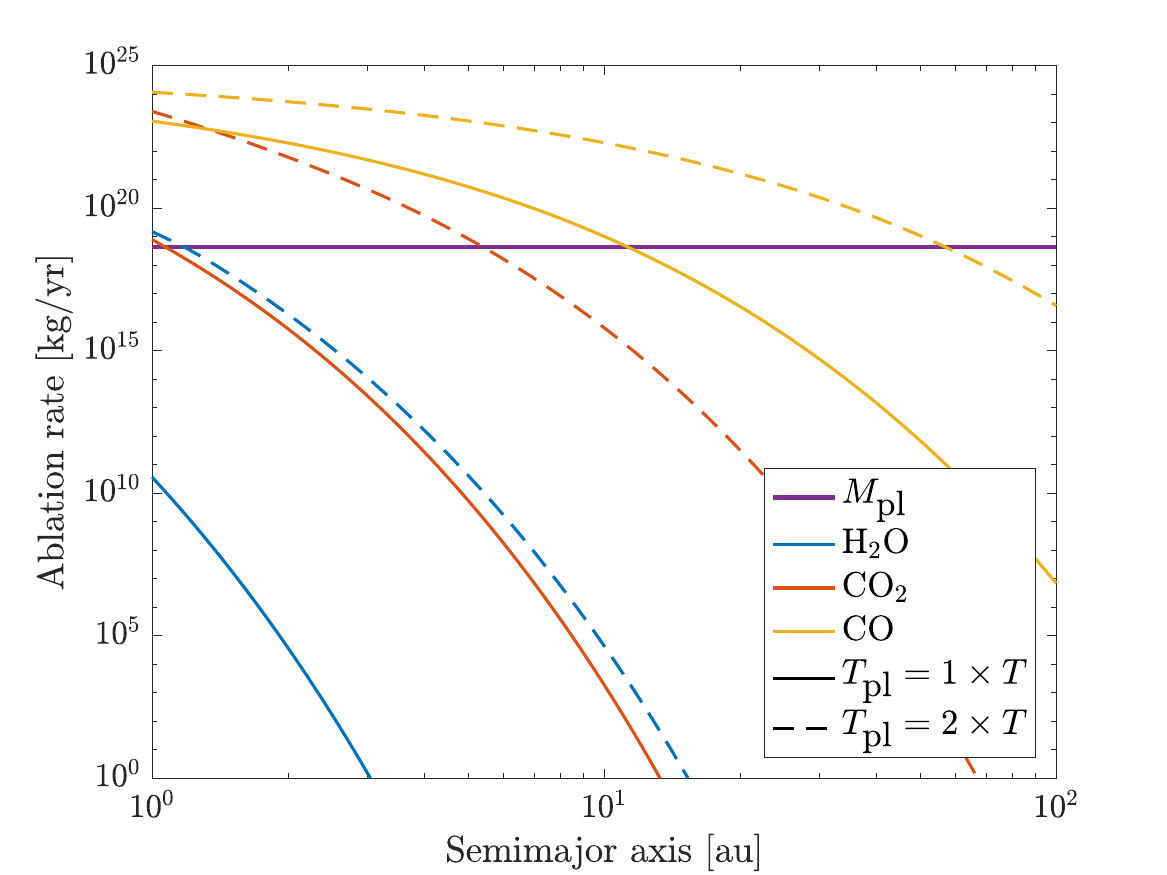}}
\caption{Mass ablation rate as a function of semimajor axis for H$_2$O, CO$_2$ and CO ice. The planetesimal surface temperature is set to equal integer multiples of the disc temperature. The mass of one planetesimal is marked as a purple line on the plot, above this line a planetesimal becomes completely ablated within one year.}
    \label{fig:ablRate_a_SS}
\end{figure}

Figure \ref{fig:ablRate_a_SS} shows the mass ablation rate as a function of the semimajor axis for H$_2$O, CO$_2$ and CO ice (calculated using equation \ref{eq:ablRate}). The ablation of H$_2$O ice is only efficient in the very innermost part of the disc, and does not generate very high ablation rates. The ablation rates for CO ice are on the contrary high enough to completely sublimate a planetesimal on a $\sim 10\, \textrm{au}$ circular orbit within a year. Planetesimals formed beyond the CO iceline should thus lose mass much further out in the disc than planetesimals formed interior of the CO iceline. The ablation rates for CO$_2$ ice lie in between the ablation rates for H$_2$O and CO ice.

\section{Varying the gap width and the planetesimal formation location}\label{Appendix: gapwidth}

\begin{figure*}
\centering
    {\includegraphics[width=.7\hsize]{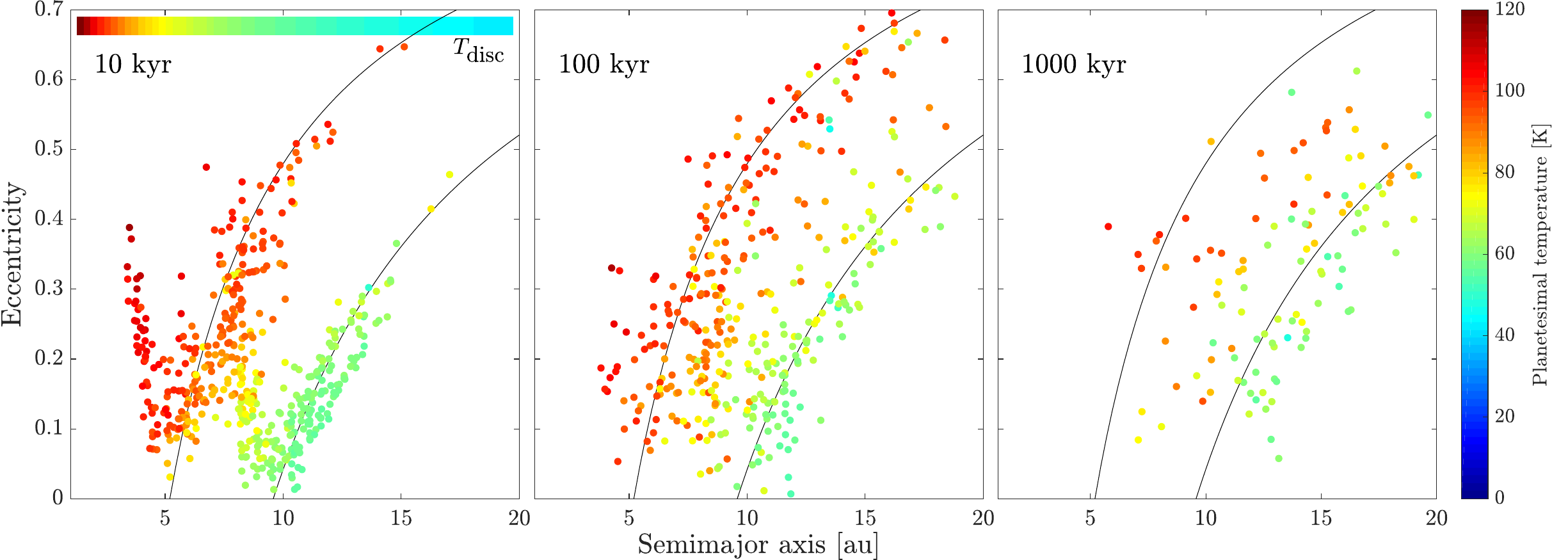}}
    {\includegraphics[width=.7\hsize]{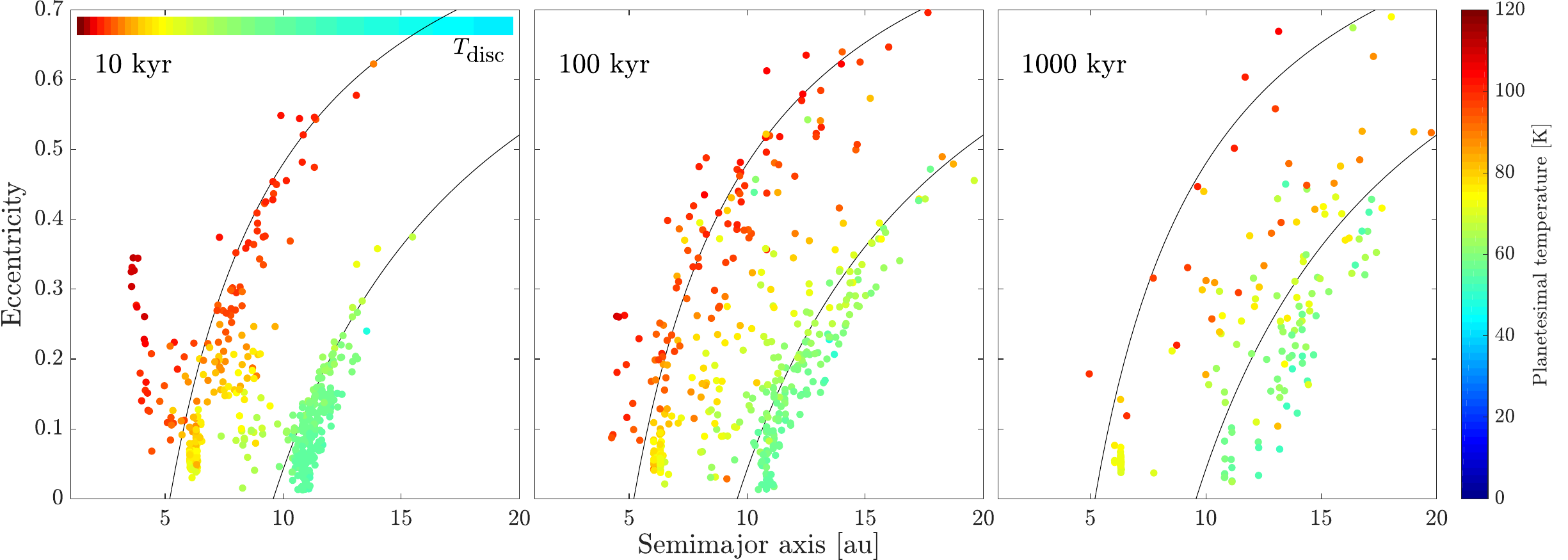}}
    {\includegraphics[width=.7\hsize]{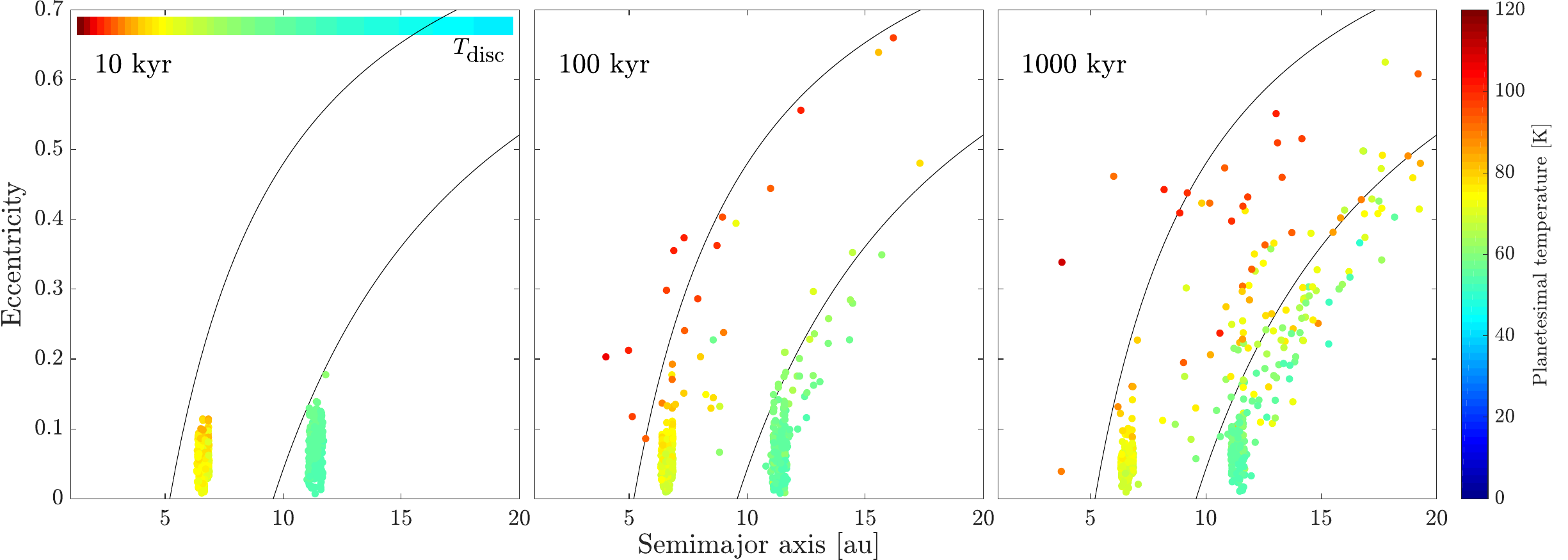}}
\caption{Eccentricity, semimajor axis and surface temperature evolution for 3 nominal Solar System simulations with 500 planetesimals per simulation and varying planetesimal formation locations. In the top panel the planetesimals are initiated between $1-3$ Hill radii away from the planets, in the middle panel between $3-5$ Hill radii, and in the bottom panel between $5-7$ Hill radii. The solid black lines mark the perihelia of the planets. Eccentricity excitation occurs much earlier and at a much faster speed for planetesimals initiated close to the planets. When the planetesimals are initiated between $5-7$ Hill radii away from the planets, a population of planetesimals still remain around their birth location at the end of the simulation. }
    \label{fig:a_e_T_nominal_rH}
\end{figure*}

\begin{figure*}
\centering
    {\includegraphics[width=\columnwidth]{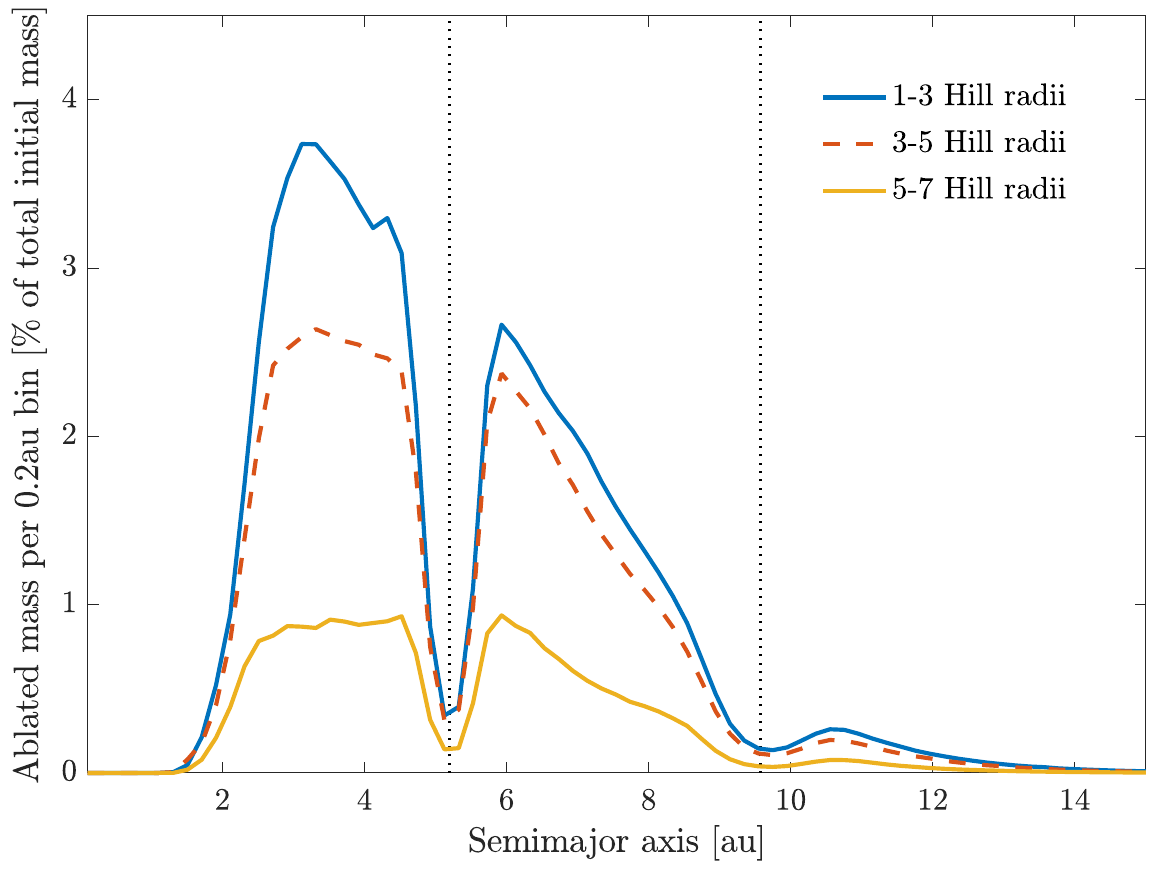}}
    {\includegraphics[width=\columnwidth]{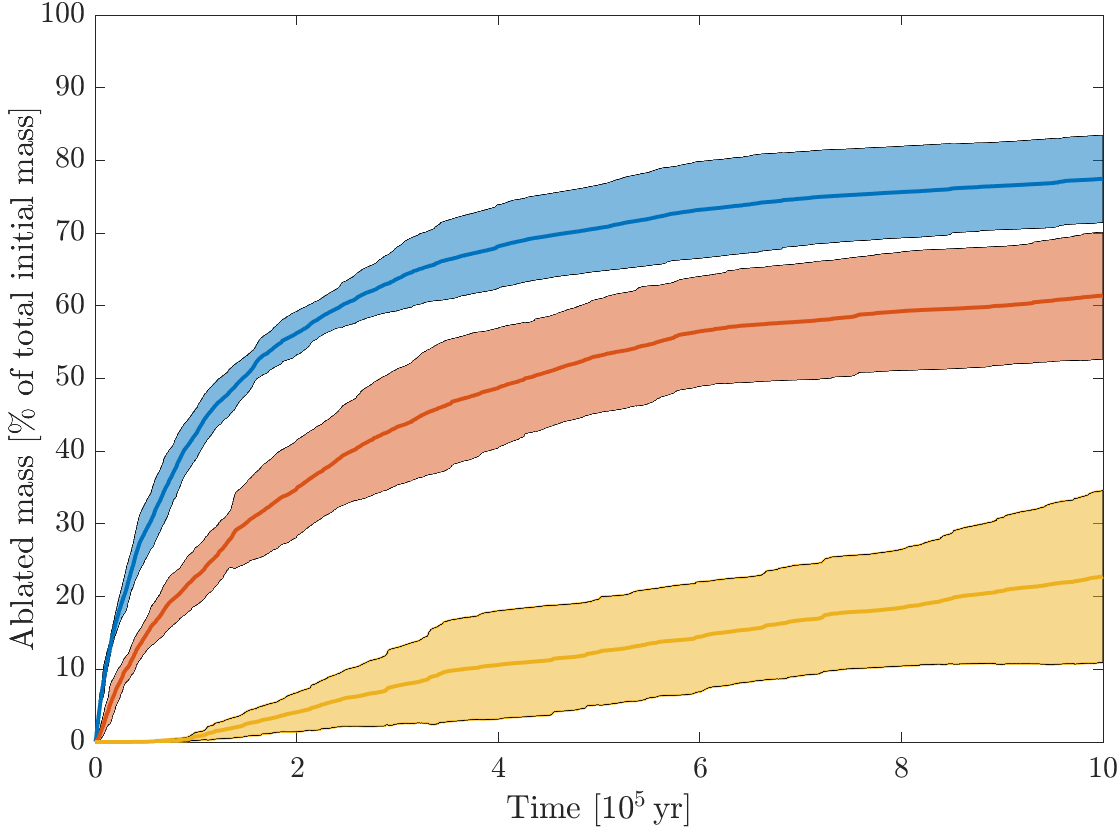}}
\caption{Left plot: distribution of ablated mass across the disc after $1\, \textrm{Myr}$ for the same 3 simulations as in Figure \ref{fig:a_e_T_nominal_rH}. The values on the $y$-axis represent the amount of mass that has been ablated in a $0.2\, \textrm{au}$ semimajor axis bin. The dotted lines mark the semimajor axes of the planets. The distribution of ablated mass is very similar in all three simulations, with the majority of mass being lost either just interior of or just exterior of Jupiter's orbit. Right plot: the total amount of mass that has been ablated as a function of time, for the same data as in the left plot. The colored lines show the average over the 5 simulations, and the colored region shows the one standard deviation away from this value. The ablation rate decreases as the planetesimal formation location w.r.t. the planet increases; however, the ablation continues for a longer period of time. }
    \label{fig:abl_planPosExp}
\end{figure*}

The planetesimals in our simulations are initiated between $1-2$ Hill radii away from the planets, which is roughly the region in which planetesimals form in Paper 1. The formation location in Paper 1 is strongly affected by (1) the width and depth of the planetary gap; and (2) our assumption that the planetesimals form at the location where the SI is triggered. Regarding (1): There are numerous gap-prescriptions in the literature, and they often disagree on the resultant gap shape. Wider gaps results in planetesimal formation further from the planet, and deeper gaps results in that the planetesimals form in a narrower region. Furthermore, 1D simulations generally produce deeper and narrower gaps than their 2D or 3D analogs. Regarding (2): \citet{Carrera2021} performed 3D simulations of the streaming instability in the presence of a pressure bump and found that the planetesimals form inward of the pressure bump that initiated their growth, so not at the exact location where the SI is triggered. Taken together, this means that we do not know exactly where the planetesimals form relative to the planet location. The formation location might be both closer or further away from the planet than it is in our simulations. 

In order to study how different formation locations affect our results we performed 3 nominal Solar System simulations where we placed the planetesimals at either $1-3$, $3-5$ or $5-7$ Hill radii away from the planets, and extended the simulation time to $1\, \textrm{Myr}$. The gas surface density profile was kept unchanged. The results of these simulations are presented in Figure \ref{fig:a_e_T_nominal_rH} and \ref{fig:abl_planPosExp}. Generally, the excitation of the orbital eccentricities is slower when the planetesimals are formed further away from the planets. Because of this, some planetesimals are still located close to their birth locations at the end of the simulation. The slower orbital excitation also results in lower ablation rates, especially in the beginning of the simulation; however, the ablation continues for a much longer period of time. This would result in that the flux of pebbles through the midplane is lower, but persists for a longer time. The presented results indicate that ablation is important also for planetesimals forming further away from the planets, but it happens on a longer timescale. 

In the above simulations we only varied the formation location of the planetesimals, not the shape of the planetary gas-gap. Narrower gaps will result in mass loss closer to the planet location, and shallower gaps will result in that more mass is lost around the planet location. We confirmed this by performing a simulation with half the gap-width and a much shallower gap. We found that the total amount of mass loss did not change, only the distribution of the ablated material around the location of the planets.

\section{Dynamical evolution of 6 planetesimals}\label{Appendix: 6plan}

\begin{figure*}
\centering
    {\includegraphics[width=\columnwidth]{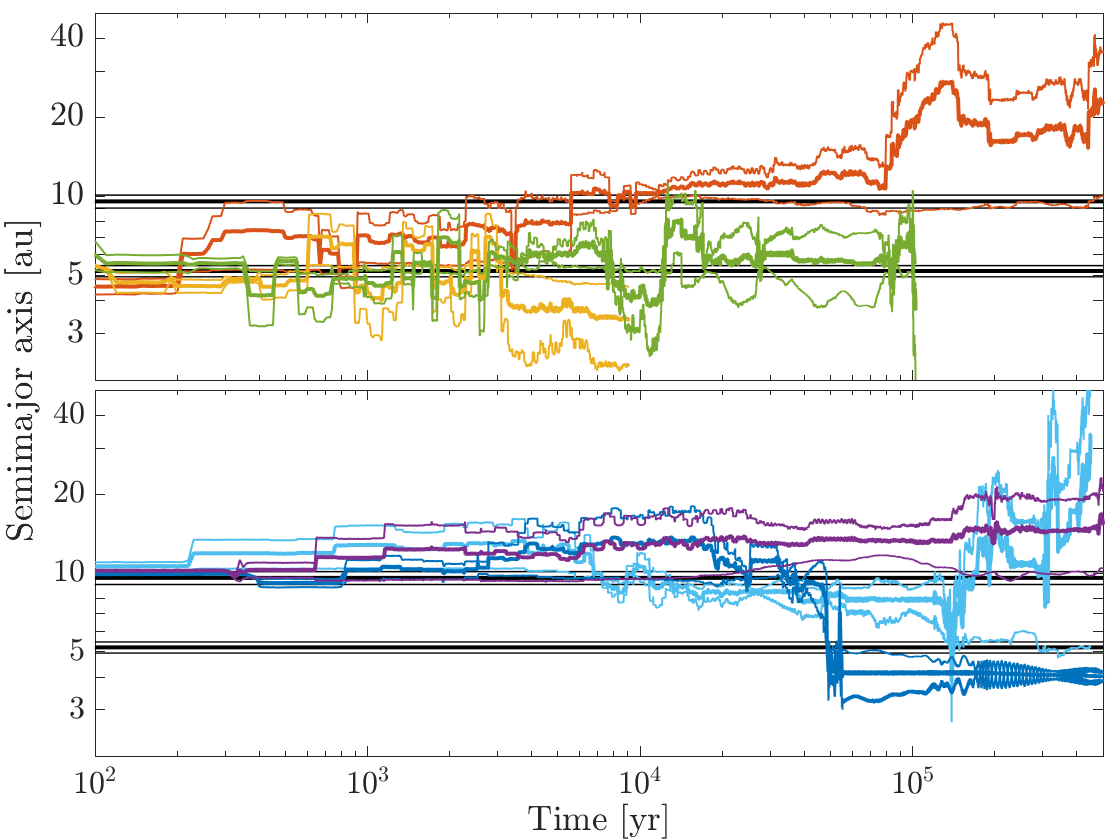}}
    {\includegraphics[width=\columnwidth]{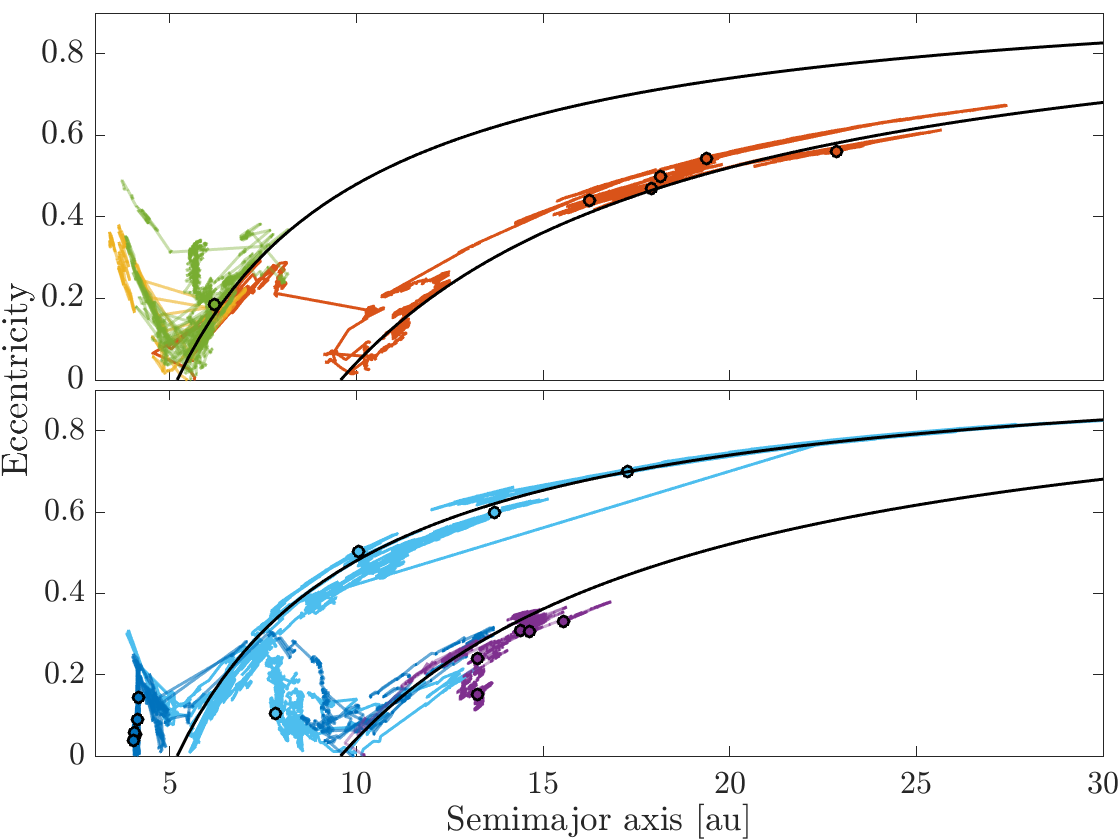}}
\caption{Semimajor axis and eccentricity evolution for the same data as in Figure \ref{fig:ml_a_6plan_nominal_SS}, with the same color-coding. Planetesimals formed at Jupiter's gap edge are plotted in the top panels, and planetesimals formed at Saturn's gap edge are plotted in the bottom panels. Left plot: Evolution of the planetesimals semimajor axis (thick colored lines), perihelion and aphelion (thin colored lines). The semimajor axes, perihelia and aphelia of Jupiter and Saturn (thick and thin black lines) are included as well. Right plot: Eccentricity and semimajor axis tracks, with dots to mark $100\, \textrm{kyr}$ of evolution. The black lines mark the perihelia of Jupiter and Saturn. }
    \label{fig:a_e_t_6plan_nom_SS}
\end{figure*}

Figure \ref{fig:a_e_t_6plan_nom_SS} shows the dynamical evolution for the same 6 planetesimals as in Figure \ref{fig:ml_a_6plan_nominal_SS}. The "red" planetesimal forms at the gap edge of Jupiter, and is a member of Jupiter's scattered disc until it obtains a strong kick and becomes a member of Saturn's scattered disc, where it remains until the end of the simulation. The "yellow" planetesimal is a member of Jupiter's scattered disc for a few thousand years, until a kick by Jupiter places it on an orbit interior of the scattered disc, where it quickly becomes ablated. The "green" planetesimal is a member of Jupiter's scattered disc during most of its lifetime, until it is kicked towards the sun and becomes ablated. 

The "light-blue" planetesimal forms at the gap edge of Saturn, and sits on an orbit just interior of Saturn's scattered disc for about $150\, \textrm{kyr}$, when it is scattered towards Jupiter's scattered disc and obtains a large eccentricity which increases with time until the planetesimal leaves the simulation domain. The "dark-blue" planetesimal is a member of Saturn's scattered disc for $50\, \textrm{kyr}$, until a strong encounter with Saturn places it on an orbit interior of Jupiter, where it becomes circularized. The "purple" planetesimal is a member of Saturn's scattered disc for most of its lifetime, except during a brief period where it has an orbit slightly outside the scattered disc.

\end{document}